\documentclass[useAMS,usenatbib]{mn2e}

\usepackage{graphicx}

\newcommand{\eq}[1]{\begin{equation}  #1 \end{equation}}

\newcommand{\eqa}[1]{\begin{eqnarray}   #1 \end{eqnarray}}

\newcommand{\br}[1]{\left( #1 \right)}
\newcommand{\bc}[1]{\left\{ #1 \right\}}
\newcommand{\bb}[1]{\left[ #1 \right]}
\newcommand{\ba}[1]{\left\langle #1 \right\rangle}

\newcommand{\nn}{\nonumber}

\newcommand{\dd}{{\rm d}}

\newcommand{\vek}[1]{\mbox{\boldmath $#1$}}

\newcommand{\refmark}[1]{\textbf{#1}}

\title[Measuring \& modelling COSMOS intrinsic galaxy ellipticities]{Intrinsic galaxy shapes and alignments I: Measuring \& modelling COSMOS intrinsic galaxy ellipticities}
\author[B. Joachimi et al.]{B.~Joachimi,$^1$\thanks{E-mail: bj@roe.ac.uk} E.~Semboloni,$^2$ P.E.~Bett,$^3$ J.~Hartlap,$^3$ S.~Hilbert,$^{4,5}$ H.~Hoekstra,$^2$ \newauthor{P.~Schneider,$^3$ and T.~Schrabback$^{3,4}$}\\
  $^1$Institute for Astronomy, University of Edinburgh, Royal Observatory, Blackford Hill, Edinburgh, EH9 3HJ, U.K.\\
  $^2$Leiden Observatory, Leiden University, P.O. Box 9513, 2300 RA, The Netherlands\\
  $^3$Argelander-Institut f\"ur Astronomie, Universit\"at Bonn, Auf dem H\"ugel 71, 53121 Bonn, Germany\\
  $^4$Kavli Institute of Particle Astrophysics and Cosmology (KIPAC), Stanford University, 452 Lomita Mall, Stanford, CA 94305, \&\\ SLAC National Accelerator Laboratory, 2575 Sand Hill Road, M/S 29, Menlo Park, CA 94025\\
  $^5$Max-Planck-Institut f{\"u}r Astrophysik, Karl-Schwarzschild-Stra{\ss}e 1, 85741 Garching, Germany
}
\date{Accepted . Received ; in original form }

\pagerange{\pageref{firstpage}--\pageref{lastpage}} \pubyear{2012}

\begin{document}
\label{firstpage}

\maketitle

\begin{abstract}
The statistical properties of the ellipticities of galaxy images depend on how galaxies form and evolve, and therefore constrain models of galaxy morphology, which are key to the removal of the intrinsic alignment contamination of cosmological weak lensing surveys, as well as to the calibration of weak lensing shape measurements. We construct such models based on the halo properties of the Millennium Simulation and confront them with a sample of $90,000$ galaxies from the COSMOS Survey, covering three decades in luminosity and redshifts out to $z=2$. The ellipticity measurements are corrected for effects of point spread function smearing, spurious image distortions, and measurement noise. Dividing galaxies into early, late, and irregular types, we find that early-type galaxies have up to a factor of two lower intrinsic ellipticity dispersion than late-type galaxies. None of the samples shows evidence for redshift evolution, while the ellipticity dispersion for late-type galaxies scales strongly with absolute magnitude at the bright end. The simulation-based models reproduce the main characteristics of the intrinsic ellipticity distributions although which model fares best depends on the selection criteria of the galaxy sample. We observe fewer close-to-circular late-type galaxy images in COSMOS than expected for a sample of randomly oriented circular thick disks and discuss possible explanations for this deficit.
\end{abstract}


\section{Introduction}
\label{sec:intro}

In the paradigm of hierarchical structure formation tidal gravitational torques and shear forces play a central role in determining the morphology and angular momenta of dark matter haloes over time. These properties affect the way galaxies form, evolve and interact with the environment. In particular, they strongly impact on the distribution of, as well as the correlations between, shapes of the observable, luminous parts of galaxies.

Consequently, the intrinsic shapes and alignments of galaxies play a dual role in cosmology: On the one hand they constitute a potentially valuable and complementary probe of galaxy formation and evolution scenarios, particularly of the influence of the large-scale gravitational potential in the galaxy's environment. On the other hand, due to the scatter and the induced intrinsic correlations, the intrinsic shape properties of galaxies feature prominently in the statistical and systematic error budgets of large-scale weak gravitational lensing surveys, thereby limiting the accuracy obtainable on dark matter, dark energy, or modified gravity constraints.

Both aspects call for a better understanding of the distributions and correlations of galaxy shapes, and their dependence on time, luminosity, environment, merger history, and other properties. The large-scale shape correlations \citep[e.g.][]{mandelbaum11,joachimi11}, the alignment of satellite galaxies on small-scales \citep[e.g.][]{hao11,hung11}, and the distribution of galaxy ellipticities \citep[e.g.][]{leauthaud07} have hitherto been studied separately. However, a successful model of intrinsic galaxy shape statistics has to explain these observations simultaneously. This paper is the first of a suite in which we make a first attempt at constructing such comprehensive models and confronting them with several sets of new and existing observational data, concentrating in this first part on one-point statistics of shapes.

A major goal of this investigation is to establish a new, complementary approach to pin down viable models of galaxy intrinsic alignments, which, besides reproducing the two-point statistics, also have to be capable of predicting the distribution of galaxy shapes among various galaxy populations. Additional constraints would be most valuable because current intrinsic alignment constraints are limited to $z \la 0.7$, do not extend to deeply non-linear scales (limited by galaxy bias measurements), and are still affected by large statistical uncertainties (\citealp{mandelbaum11,joachimi11}, and references therein).

This is particularly problematic for upcoming weak lensing surveys like KiDS\footnote{\texttt{http://www.astro-wise.org/projects/KIDS}}, DES\footnote{\texttt{http://www.darkenergysurvey.org}}, HSC\footnote{\texttt{http://www.naoj.org/Projects/HSC/index.html}}, LSST\footnote{\texttt{http://www.lsst.org}}, WFIRST\footnote{\texttt{http://jdem.gsfc.nasa.gov/}}, and \textit{Euclid}\footnote{\texttt{http://www.euclid-ec.org}; \citet{laureijs11}}, which will have a large fraction if not the bulk of their source galaxies at $z>0.7$ and retrieve the majority of their potentially excellent constraints on cosmology from the non-linear regime of structure formation \citep[e.g.][]{takada04,laureijs11}. Analytical work, N-body simulations, and observations agree that intrinsic alignments could constitute a contamination of the order $10\,\%$ to weak lensing two-point statistics \citep[e.g.][]{heavens00,catelan01,heymans06,mandelbaum06}.

Hence methods designed to remove or calibrate the intrinsic alignment signal are a necessity but currently have to work under minimal assumptions about the form of the intrinsic correlations \citep[see e.g.][]{king02,king03,bridle07,joachimi08b,joachimi09,bernstein08,joachimi10,zhang10}. A robust prediction of the intrinsic alignment contamination, which this work aims at, will therefore be a reliable base for developing weak lensing survey strategies and tools to control intrinsic alignments.

Much of the early work on the statistical properties of galaxy morphologies has focused on inferring the three-dimensional shapes of galaxies from their light distributions \citep{binggeli80,binney81}. \citet{lambas92} analysed axis ratios of images in the APM Bright Galaxy Survey and found significant differences in the frequency of small axis ratios between their early- and late-type samples. 

The distribution of galaxy ellipticities in fainter samples has primarily been investigated to assess statistical error limits on weak lensing measurements \citep[e.g.][]{brainerd96,bernstein02,leauthaud07}. Nonetheless, some of these results also provided hints at clear differences in the ellipticity distributions between different galaxy populations \citep[e.g.][]{uitert12}, and constrained the evolution of the dispersion of intrinsic ellipticities with redshift \citep{leauthaud07}, indicating that these measures may add considerable constraining power on galaxy shape models.

Substantial differences in the dispersion of intrinsic ellipticities between different galaxy populations could have interesting implications for measurements of large-scale weak gravitational lensing. Forthcoming surveys will cover large areas of the sky and will thus be limited by the ellipticity noise on medium and small scales from which most of the cosmological information is extracted. The noise power spectrum is proportional to $\sigma_\epsilon^2/n_{\rm g}$, where $\sigma_\epsilon$ is the dispersion of the complex ellipticity and $n_{\rm g}$ the projected number density of galaxies with shape measurements \citep{bartelmann01}. 

Therefore certain galaxy samples, appropriately selected to have low ellipticity dispersion, can beat down statistical error limits or become a valid alternative despite lower number density. Using such samples might be desirable if e.g. shape measurements became easier, photometric redshifts more precise, or intrinsic alignments of galaxy shapes either intrinsically weaker or easier to pin down with external data.

Recent progress in the gravitational shear estimation from galaxy images \citep[e.g.][]{kitching12,refregier12} demonstrates the importance of biases introduced by noise in the images. As these biases depend on galaxy ellipticity \citep{melchior12}, it is paramount to know the distribution of intrinsic galaxy ellipticities of the sample under consideration. Intrinsic ellipticity distributions with negligible measurement noise contributions are challenging to determine observationally, so that the ability to reliably model the intrinsic shapes of arbitrary galaxy samples is most desirable in the light of forthcoming weak lensing surveys.

In the following we will extract intrinsic ellipticity dispersions and distributions of ellipticities from the \textit{HST} COSMOS Survey \citep{scoville07} and confront these measurements with simulation-based models, with the aim of interpreting the statistical properties of galaxy shapes in COSMOS, identifying samples that could reduce the noise limits of weak lensing surveys, and select realistic models of galaxy morphology. In a forthcoming paper we will then use the same models to investigate intrinsic shape correlations, match them against current observational constraints, and predict the intrinsic alignment contamination on planned weak lensing surveys.

As currently it is computationally not yet possible to run high-resolution hydrodynamic simulations on a cosmological volume, we will rely on a dark matter-only simulation complemented with a \lq semi-analytic\rq\ model of the galaxy morphology. Our galaxy shape models are based on the halo properties extracted from the Millennium Simulation \citep{springel05}, which comprises a sufficiently large volume to allow for a measurement of large-scale correlations, but also has excellent mass resolution (see e.g. \citealp{heymans06} whose simulations have 20 times higher particle mass).

Correlations of dark matter halo ellipticities and angular momenta among each other and with the large-scale matter distribution have been investigated in great detail with N-body simulations \citep[e.g.][]{bailin05,altay06,hahn07b,lee08}. We will supplement this information with multi-band photometry and galaxy type classifications from the semi-analytic models of galaxy formation and evolution by \citet{bower06}, which enables an accurate selection of galaxy samples for comparison with observations.

As we rely on a dark matter-only simulation, our galaxy shape models have to make assumptions about how baryons trace the dark matter. We will follow earlier simulation-based work \citep{heavens00,heymans06} and analytic intrinsic alignment models \citep[see e.g.][]{catelan01,hirata04} in assuming that early-type galaxies have the same shapes as their dark matter haloes, and that late-type galaxies are composed of thick disks perpendicular to the angular momentum of the halo.

A large number of small-scale, high-resolution hydrodynamic simulations \citep{bosch02,croft09,hahn10,bett10,bett11} have been analysed to yield statistical properties of the relation between luminous and dark matter, which we incorporate into the models. Moreover the Millennium data includes the positions of satellite galaxies, but no shapes as the corresponding subhaloes are not sufficiently resolved. Hence we resort to simple models of satellite shapes (and alignments), partly based on the high-resolution simulations by \citet{knebe08}; see also \citet{kuhlen07,pereira08,faltenbacher08,knebe10} for similar investigations into the shapes of satellite galaxies and halo substructure.

This article is organised as follows. In Section \ref{sec:sims} we summarise the main aspects of the underlying simulations and the quantities derived therefrom, before detailing in Section \ref{sec:galaxymodels} the modelling of galaxy shapes. We provide an overview on the extraction and processing of intrinsic galaxy ellipticities from the COSMOS Survey in Section \ref{sec:cosmos}. In Section \ref{sec:cosmosresults} we present the results of our observational analysis and compare them with various galaxy shape models. We summarise and conclude on our findings in Section \ref{sec:conclusions}.

Unless stated otherwise, rest-frame magnitudes are $k+e$-corrected to $z=0$ and computed assuming the cosmology of the Millennium Simulation (see below) except for a Hubble constant $H_0=100\,h {\rm km/s/Mpc}$ with $h=1$. Magnitudes extracted from the Millennium data base are given in the Vega system, while all observations use the AB system. If direct comparison is necessary, we resort to the conversion tables of \citet{fukugita96}.

\section{Simulations}
\label{sec:sims}

\subsection{N-Body simulation}
\label{sec:nbody}

As the basis for our galaxy models we require the shapes and angular momenta of the underlying dark matter distribution, which we obtain from the Millennium Simulation \citep{springel05}. With a comoving box size of $500\,{\rm Mpc}/h$ populated with $2160^3$ particles of mass $m_{\rm p}=8.6 \times 10^8\,h^{-1} M_\odot$, the Millennium Simulation provides us with a representative sample of the Universe with the resolution necessary to determine the properties of galaxy-sized dark matter haloes accurately. 

The simulation followed the evolution of the matter distribution with 64 snapshots from $z=127$ to $z=0$ using the TreePM algorithm of \texttt{GADGET-2} \citep{springel05b} with a comoving force softening scale of $5\,{\rm kpc}/h$. The underlying cosmology is a spatially flat $\Lambda$CDM universe with matter density parameter $\Omega_{\rm m}=1-\Omega_\Lambda=0.25$ at redshift zero. The $z=0$ baryon density parameter is $\Omega_{\rm b}=0.045$, the Hubble parameter $h=0.73$, the power-law index of the initial power spectrum $n_{\rm s}=1$, and the normalisation of the power spectrum $\sigma_8=0.9$.

These parameters were chosen to be consistent with results from the 2dF redshift survey \citep{percival02} and the 1st year data of WMAP \citep{spergel03}. More recent analyses however suggest a significantly smaller value of $\sigma_8$ around 0.8 \citep{komatsu10,schrabback09}. The impact of such a change in the normalisation of matter fluctuations on the shapes and alignments of dark matter halo shapes and angular momenta is not yet well understood, but might become particularly relevant on small scales where non-linear gravitational physics dominates. 

\citet{allgood06} measured the length ratios of the smallest to largest eigenvector of the halo mass distributions in simulations with $\sigma_8=0.9$ and $\sigma_8=0.75$, finding that higher $\sigma_8$ creates on average more spherical haloes. This could be due to more non-linear evolution and more frequent mergers or due to an earlier halo collapse when the Universe still had a smoother matter distribution. From their Fig.$\,$3 we estimate that a change in $\sigma_8$ from 0.9 to 0.75 modifies the mean of the projected ellipticity of haloes by not more than 0.05 for $0 \leq z \leq 1$. Baryonic physics in the central region of haloes also tends to decrease halo ellipticity \citep{kazantzidis04} although the magnitude of this effect is still uncertain. We conclude that the impact of the high value of $\sigma_8$ in the Millennium Simulation is small for the purposes of this pilot study, and besides it mimics to some extent expected baryonic effects.

Ray-tracing through the Millennium Simulation was performed by \citet{hilbert09}. We will use those catalogues which were constructed from 64 light cones with an area of $4 \times 4\,{\rm deg}^2$ each (note that we do not require the gravitational shear measurements in this work). This results in a mock survey of $1024\,{\rm deg}^2$ out to a redshift of $z \approx 2.1$. After imposing a magnitude limit of $F814W<24$ (obtained via the semi-analytic models; see Sections \ref{sec:semianalytics} and \ref{sec:cosmosmethod} for details), the survey has a mean galaxy number density of about $30\,{\rm arcmin}^{-2}$.

\subsection{Halo shapes and angular momenta}
\label{sec:shapes}

\begin{figure}
\centering
\includegraphics[scale=.33,angle=270]{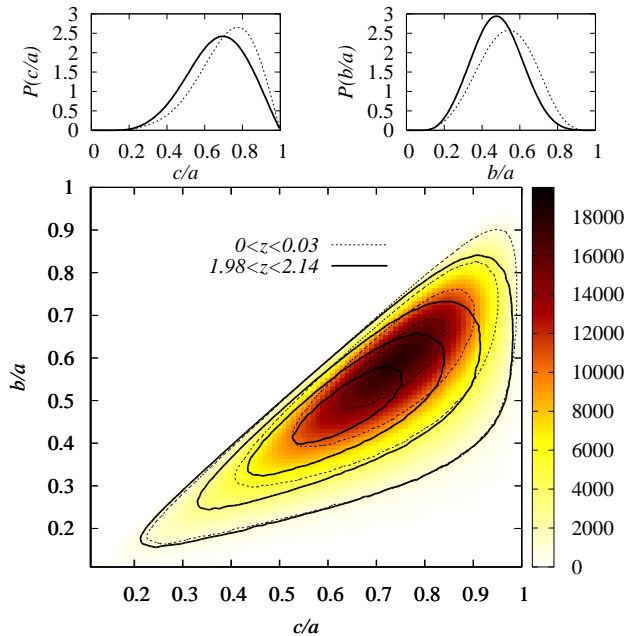}
\caption{Distribution of halo shapes in the Millennium simulation. \textit{Bottom panel}: Number of haloes as a function of their axis ratios (where the axis lengths $c \leq b \leq a$ are the square roots of the eigenvalues of the halo inertia tensor), in the redshift ranges $0 < z < 0.03$ (dotted contours and colour scale) and $1.98 < z < 2.14$ (solid contours). Contours are plotted at $1000$, $5000$, $10000$, and $15000$ haloes. Note that in this plot prolate haloes reside along the diagonal, oblate haloes along the right margin, and spherical haloes in the upper right corner. \textit{Top panels}: Probability density of the axis ratios $c/a$ and $b/a$ for haloes with $0 < z < 0.03$ (dotted lines) and $1.98 < z < 2.14$ (solid lines)}.
\label{fig:shapedistribution}
\end{figure} 

We follow \citet{bett07} in identifying bound structures in the simulation and in computing their shape and angular momenta. A dark matter halo is defined as a collection of self-bound sub-haloes, i.e. single unbound particles get discarded. Firstly, groups of simulation particles were constructed with a friends-of-friends algorithm \citep{davis85}, followed by the identification of subhaloes as self-bound structures within these groups via \texttt{SUBFIND} \citep{springel05}. Merger-tree data is then used to identify and remove subhaloes that are only transiently in proximity to the halo (these are then treated as separate haloes). This procedure removes many of the problems associated with friends-of-friends halo identification without biasing the halo shape towards a spherical boundary, as discussed in \citet{bett07}. The resulting halo definition corresponds to the \lq merger-tree\rq\ haloes described by \citet{harker06}.

The \lq shape\rq\ of each halo is computed via the quadrupole tensor of the mass distribution per unit mass $\mathbf{M}$ with components
\eq{
\label{eq:massquadrupole}
M_{\mu\nu} = \sum_{i=1}^{N_{\rm p}} r_{i,\mu} r_{i,\nu}\;,
}
where $N_{\rm p}$ is the number of particles in the halo, and where $\vek{r}_i$ denotes the position vector of particle $i$ with respect to the halo centre (defined as the location of the gravitational potential minimum). The eigenvalues and eigenvectors of $\mathbf{M}$ define an ellipsoid, with the eigenvalues per unit mass giving the square semi-axis lengths $c^2 \leq b^2 \leq a^2$, and the corresponding eigenvectors specifying the axis orientations. We interpret this ellipsoid as an approximation to the shape of the halo. In Fig.$\,$\ref{fig:shapedistribution} we have plotted histograms of the two axis ratios of the resulting ellipsoids for redshifts around 0 and 2. Haloes are preferentially prolate and tend to be closer to spherical at low redshift.

\citet{bett07} recommended a minimum particle number of $N_{\rm p}=300$ to avoid biases in shape measurement. Since this a restrictive condition that would discard more than half of the haloes identified in the Millennium Simulation, we assess whether we can decrease this threshold, measuring the accuracy of halo shape as a function of $N_{\rm p}$. To this end we create mock ellipsoidal haloes with a radial NFW mass profile and randomly populated with $N_{\rm p}$ equal-mass particles. We assume a concentration of 10 and truncate the halo at the virial radius. The shapes of these mock haloes are then measured via the method outlined above, using 100 haloes for each value of $N_{\rm p}$ that we test. We vary $N_{\rm p}$ from $10$ to $10000$ in 12 approximately logarithmic steps and take input axis ratios in the range $0.1 \leq c/a, b/a \leq 0.9$. 

\begin{figure*}
\centering
\includegraphics[scale=.35,angle=270]{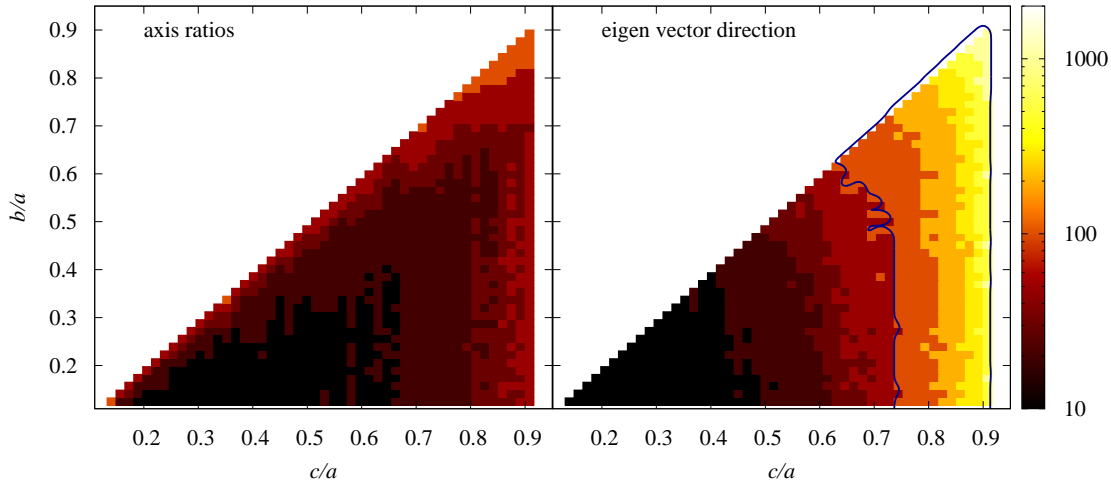}
\caption{Minimum number of particles per halo required for accurate shape measurement. \textit{Left panel}: Number of halo particles needed to achieve a $5\,\%$ maximum deviation of the measured axis ratios of the ellipsoid from the input values. The required number is less than 300 throughout, except for the far top right corner. \textit{Right panel}: Number of halo particles needed to achieve a $5^\circ$ maximum deviation of the measured orientation of the largest halo eigenvector from the input value. The blue contour indicates the region where the required particle number exceeds 300. Note however that for oblate haloes the directions of the two largest eigenvectors become degenerate, so that a deviation larger than $5^\circ$ is tolerable.}
\label{fig:nfwhalosample}
\end{figure*} 

In Fig.$\,$\ref{fig:nfwhalosample} we show for every combination of $c/a$ and $b/a$ the number of halo particles required to achieve less than $5\,\%$ deviation of the measured axis ratios from their input values and less than $5^\circ$ deviation of the measured orientation of the largest halo eigenvector from the input direction. While generally a few tens of particles are sufficient to measure axis ratios for triaxial haloes, the requirements on particle number become more stringent if two of the axes are of similar size, i.e. for strongly prolate and oblate haloes. 

The threshold of 300 particles is only exceeded for close to spherical halo shapes, which are not important for our analysis as the number of haloes with this shape are small (see Fig.$\,$\ref{fig:shapedistribution}), and because these haloes have very low ellipticity in projection on the sky. The accuracy for the eigenvector direction is not met with 300 halo particles or less for oblate haloes with $c/a \ga 0.7$, but note that in this case the directions of the two largest eigenvectors become degenerate, so that a deviation larger than $5^\circ$ is acceptable for our purposes. The latter result is in good agreement with the limit of $c/a = 0.81$ found for haloes with $N_{\rm p}=1000$ particles by \citet{bett10}.

The quantity which will eventually be used for further analysis is the projected ellipticity of the halo in Cartesian coordinates (see Section \ref{sec:galaxymodels} for details). Placing the largest eigenvector along the line of sight, we compute the accuracy in the ellipticity components attainable with 300 halo particles. The difference between actual and recovered ellipticity varies only weakly with the values of the axis ratios $c/a$ and $b/a$, between around 0.01 for haloes that are strongly elliptical in projection and around 0.02 for haloes with nearly spherical projections.

When accepting $10\,\%$ deviation of the measured axis ratios and $10^\circ$ deviation of the measured orientation of the largest eigenvector, we find that the minimum requirement can be relaxed to $N_{\rm p}=100$ with similar accuracy as shown in Fig.$\,$\ref{fig:nfwhalosample}. We will adopt this less stringent limit when modelling the shapes of central early-type galaxies.

These lower limits on particle numbers for shape computation are of the same order as those deduced by \citet{jing02} and \citet{pereira08}, the latter paper presenting a similar approach based on mock NFW haloes for the reduced inertia tensor $M^{\rm red}_{\mu\nu} = \sum_{i=1}^{N_{\rm p}} r_{i,\mu} r_{i,\nu} / r_i^2$. \citet{jing02} derive a minimum number of 160 for which ellipticity correlations are underestimated not more than $5\,\%$ with respect to high-resolution simulations\footnote{Note however that even with several hundreds of particles halo shape measurements can still be afflicted with resolution issues, e.g. due to unresolved substructure that affects the definition of subhaloes via \texttt{SUBFIND} \citep{schneiderm11}.}. We agree with this work in that sparsely sampled haloes tend to produce smaller axis ratios (and hence larger ellipticity on average) as well as rapidly increasing uncertainty in the halo orientation. The net effect is an underestimation of ellipticity correlations by up to a factor of 2 for haloes with 20 particles \citep{jing02}.

We calculate the specific angular momentum of haloes,
\eq{
\label{eq:specificangmom}
\vek{L} = \frac{1}{N_{\rm p}(r_{\rm vir})} \sum_{i=1}^{N_{\rm p}(r_{\rm vir})} \vek{r}_i \times \vek{v}_i\;,
}
where $\vek{v}_i$ is the velocity of particle $i$ relative to the halo centre of mass velocity. Only particles within the virial radius are included in the sum. \citet{bett07} investigate the minimum number of particles needed for accurate angular momentum calculations by comparing with a low-resolution version of the Millennium Simulation. As their Fig.$\,$7 demonstrates, below $N_{\rm p} \approx 300$ the limited resolution of the halo causes a sharp upturn in the spin parameter; therefore, we adopt this threshold for our computations.

\subsection{Semi-analytic models}
\label{sec:semianalytics}

We supplement the information extracted from the simulation with apparent and rest-frame magnitudes in various bands of galaxies hosted by the dark matter haloes using the semi-analytic galaxy evolution model \texttt{GALFORM} in the version of \citet{bower06}. Its main updates on previous implementations concern the explicit tracking of AGN evolution and feedback, the improved modelling of disk instabilities and gas cooling, as well as the use of the merger trees by \citet{harker06}; for details see \citet{bower06} and references therein.

\citet{parry09} classified galaxy morphologies via the bulge-to-total ratio of rest-frame $K$-band luminosity, $R_{\rm type}=L_{K, {\rm bulge}}/L_{K, {\rm total}}$, where the $K$-band closely follows stellar mass over a wide range of redshifts and is robust to uncertainties in modelling details such as reddening. Defining spiral galaxies via $R_{\rm type}<0.4$, S0 galaxies via $0.4 < R_{\rm type} < 0.6$, and elliptical galaxies via $R_{\rm type}>0.6$, the \citet{bower06} models yield a distribution of morphologies at low redshift that is consistent with observations.

The merger histories show a clear dichotomoy between ellipticals on the one hand and S0 and spiral galaxies on the other hand, in particular with respect to the fraction of major merger events \citep{parry09}. Since the merger history is thought to be decisive for how the morphology of the bright part of a galaxy is related to halo properties, we use the threshold $R_{\rm type}=0.6$ to discriminate between our early-type and late-type galaxy shape models (see below). This classification can also be motivated intuitively: Lenticular galaxies are disk-dominated systems; their shape is thus thought to be determined by angular momentum, similar to spiral galaxies.

The evolutionary models also keep track of whether a galaxy is \lq central\rq, defined as the galaxy in the most massive substructure of a halo at any given time. All other galaxies in the halo are \lq satellites\rq, and are treated differently with regard to e.g. gas accretion/stripping and orbits (see \citealp{cole00} for details). We adopt this distinction in the modelling of galaxy shapes.

\section{Galaxy shape modelling}
\label{sec:galaxymodels}

Our modelling of galaxy shapes adopts the scheme of \citet{heymans06} in dividing a galaxy sample into late types whose shapes are determined by the angular momentum of the underlying dark matter halo and early types whose shapes follow the shape of their haloes. The shapes based on halo properties are assigned to the galaxies identified by the semi-analytic models as central to the halo, while we sample satellite shapes and orientations from distributions extracted from the Millennium and other simulation works. An overview on the different models presented and explored in the following is given in Table \ref{tab:galaxymodels}.

Note that, while the choice of orientation has a strong impact on the projected shape of an individual galaxy, it is irrelevant for the probability distribution of ellipticities of an ensemble of galaxies in an isotropic universe.

\begin{table*}
\centering
\caption{Overview on models for galaxy shapes. The distinction between central and satellite galaxy is adopted from the semi-analytic model. \lq Early-type\rq\ galaxies have $R_{\rm type}>0.6$, \lq late-type\rq\ galaxies $R_{\rm type}<0.6$. The rightmost column contains the identifiers used to construct the names of shape models. Note that low-mass galaxies with too few particles in their haloes to make accurate shape ($N_p<100$) and angular momentum ($N_p<300$) measurements are assigned random orientations, but otherwise follow the model assumed for the respective central galaxy type.}
\begin{tabular}[t]{llll}
\hline\hline
halo type & galaxy type & model & identifier\\
\hline
 & early-type & same shape as halo; simple inertia tensor & \texttt{Est}\\
central & " & same shape as halo; reduced inertia tensor & \texttt{Ert}\\
 & late-type & thick disk $\perp$ angular momentum of halo; $r_{\rm edge-on}=0.25$ & \texttt{Sma}\\
 & " & thick disk $\perp$ angular momentum of halo; $r_{\rm edge-on}=0.1$ & \texttt{Sth}\\
\hline
 & early-type & shape sampled from MS halo distribution; simple inertia tensor & \texttt{est}\\
 & " & shape sampled from MS halo distribution; reduced inertia tensor & \texttt{ert}\\
satellite & " & \citet{knebe08} shape modifications & \texttt{ekn}\\
 & late-type & thick disk, $r_{\rm edge-on}=0.25$ & \texttt{sma}\\
 & " & thick disk, $r_{\rm edge-on}=0.1$ & \texttt{sth}\\
\hline
\end{tabular}
\label{tab:galaxymodels}
\end{table*}

\subsection{Early-type galaxies}
\label{sec:Emodels}

All central galaxies with $R_{\rm type} \geq 0.6$ and in haloes with more than 100 particles are assumed to have the same three-dimensional shape as their host haloes. More precisely, we project the ellipsoid defined by the eigenvectors and eigenvalues of the halo inertia tensor onto the plane of the sky and treat the resulting ellipse as the shape of the galaxy. 

Let the three unit eigenvectors of the halo inertia tensor be denoted as $\vek{s}_\mu=\bc{s_{x,\mu},s_{y,\mu},s_{\parallel,\mu}}^\tau$ and the absolute values of the semi-axes as $\omega_\mu$ for $\mu=1,2,3$. Then the projected ellipse is given by all points $\vek{x}$ in the plane of the sky which fulfil $\vek{x}^\tau {\mathbf W}^{-1} \vek{x} = 1$, where we have defined
\eq{
\label{eq:ellipseprojection1}
{\mathbf W}^{-1} = \sum_{\mu=1}^3 \frac{\vek{s}_{\perp,\mu} \vek{s}_{\perp,\mu}^\tau}{\omega_\mu^2} - \frac{\vek{k} \vek{k}^\tau}{\alpha^2}\;,
}
using
\eq{
\label{eq:ellipseprojection2}
\vek{k} = \sum_{\mu=1}^3 \frac{s_{\parallel,\mu} \vek{s}_{\perp,\mu}}{\omega_\mu^2}\;~~~\mbox{and}~~
\alpha^2 = \sum_{\mu=1}^3 \br{ \frac{s_{\parallel,\mu}}{\omega_\mu} }^2\;.
}
Here, $\vek{s}_{\perp,\mu} = \bc{s_{x,\mu},s_{y,\mu}}^\tau$ corresponds to the eigenvector projected along the line of sight. A detailed derivation of the foregoing equations is provided in Appendix \ref{app:ellipse}.

The galaxy ellipticity is then defined in terms of the complex polarisation $e$ (see \citealp{bartelmann01} for details and other ellipticity definitions), computed from the symmetric tensor ${\mathbf W}$ via
\eqa{
\label{eq:ellipseprojection3}
e_1 &=& \frac{W_{11}-W_{22}}{W_{11}+W_{22}}\;;\\ \nn
e_2 &=& \frac{2\, W_{12}}{W_{11}+W_{22}}\;.
}
Note that in the special case that the shortest eigenvector lies along the line of sight, the absolute value of the polarisation is given by $|e|=(a^2-b^2)/(a^2+b^2)$. The projection implicitly assumes that the three-dimensional light distribution is uniform with a sharp cut-off at the edges. We refrain from using more complicated schemes involving a realistic radial light distribution as this could imply variable ellipticity as a function of radius, e.g. manifested as isophote twisting.

We decide to use the shape of the full halo to model the galaxy because haloes by definition should be virialised and thus would ideally have well-defined, stable shapes. Substructures, including the most massive ones that host central galaxies, are gravitationally bound but not required to be in equilibrium, and thus do not necessarily have as well-defined boundaries or as stable shapes. Moreover, unresolved luminous substructure contributes to the ellipticity of the light distribution of a galaxy, so it seems reasonable that the corresponding substructure in the underlying matter distribution is taken into account (at least on scales much smaller than the virial radius).

However, many authors studying the morphology of dark matter haloes employ the reduced inertia tensor to determine shapes, arguing that giving more weight to the inner part of a halo reduces the influence of the distribution of subhaloes in the outskirts and produces a better approximation of the shape of the galaxy residing close to the halo centre. \citet{bett11} studied the impact of different halo shape measurement algorithms using the same data set. Figure 3 of that work demonstrates that switching from the simple inertia tensor (see Equation \ref{eq:massquadrupole}), to the reduced one increases the minor-axis to major-axis ratio by about $25\,\%$, with only a weak dependence on halo mass. Assuming that a similar modification also occurs for the intermediate-axis to major-axis ratio, we rescale all semi-axes accordingly to obtain a model based on the more spherical haloes resulting from reduced inertia tensor measurements.

For galaxies classified as early type whose haloes have $N_{\rm p} < 100$ we cannot reliably measure their shapes. Instead, we assume that the statistical halo shape properties of galaxies with $N_{\rm p} < 100$ are the same as those of more massive galaxies. In each redshift slice we construct two-dimensional histograms of halo axis ratios from haloes with $N_{\rm p} \geq 300$ like those shown in Fig.$\,$\ref{fig:shapedistribution}. The low-mass galaxies at the same redshift are then assigned halo shapes which are randomly sampled from these histograms. Around $27\,\%$ of the early-type galaxies with $F814W<24$ are modelled in this way.

\subsection{Late-type galaxies}
\label{sec:Smodels}

All central galaxies with $N_{\rm p} \geq 300$ and $R_{\rm type} < 0.6$ are modelled as circular thick disks whose orientation is determined by the angular momentum of the underlying halo. If the rotation axis of the disk is perfectly aligned with the angular momentum vector $\vek{L}=\bc{L_x,L_y,L_\parallel}^\tau$, the polarisation of the galaxy image is given by
\eqa{
\label{eq:spiraleps1}
e_1 &=& \cos (2 \theta)\; \frac{1-r^2}{1+r^2}\;;\\ \nn
e_2 &=& \sin (2 \theta)\; \frac{1-r^2}{1+r^2}\;,
}
where the polar angle of the image ellipse is computed via
\eq{
\label{eq:spiraleps2}
\theta = \frac{\pi}{2} + \arctan \br{ \frac{L_y}{L_x} }\;.
}
The axis ratio of the ellipse is readily calculated as 
\eq{
\label{eq:spiraleps3}
r = \frac{|L_\parallel|}{|\vek{L}|} + r_{\rm edge-on}\; \sqrt{1 - \frac{L_\parallel^2}{|\vek{L}|^2}  }\;,
}
where $r_{\rm edge-on}$ is the ratio of disk thickness to disk diameter, i.e. approximately the axis ratio for a galaxy viewed edge-on. We again assume a uniform light distribution in the disk with a sharp cut-off at the perimeter. Moreover we neglect any small deviations of the image from an elliptical shape in the projection.

For a disk similar to the one of the Milky Way one expects $r_{\rm edge-on}$ to be of the order 0.1, but a representative sample of late-type galaxies viewed edge-on should have significant contributions by a bulge. \citet{bailin08} plot isophotal axis ratios of a large sample of SDSS galaxies in their Fig.$\,$4, finding that the smallest ratios for late-type galaxies are indeed close to 0.1. Furthermore the distribution quickly drops off below $r_{\rm edge-on}=0.25$, which we therefore choose as an alternative value to explore\footnote{Note that \citet{bailin08} employ isophotal shape measurements which are prone to biases by noise, substructure, and blending. Moreover they do not correct for the point spread function (PSF), so that the extracted value for $r_{\rm edge-on}$ can only be considered a rough estimate.}. 

Note that if we had incorporated a spheroidal component explicitly into our models, we would again be faced with a radial ellipticity gradient across the projected galaxy images, which is beyond the scope of this work. The impact of bulges implies a distribution of isophotal axis ratios for galaxies viewed edge-on, where typical values should be bracketed by our two choices of $r_{\rm edge-on}$. \citet{heymans04,heymans06} used a similar prescription for late-type galaxy models. Disk thickness is accounted for by rescaling ellipticities as $\epsilon_{\rm gal} = 0.73\, \epsilon_{\rm thin~disk}$, which in the edge-on limit corresponds to an axis ratio of 0.16, hence lying in-between the models we consider. 

Analogous to low-mass early-type galaxies, central late-type galaxies with $N_{\rm p} < 300$ that have no angular momentum information (around $56\,\%$ of all late-type galaxies with $F814W<24$) are modelled as randomly oriented thick disks, where $r_{\rm edge-on}$ has the same value as the model used for the corresponding model of central late-type galaxies with $N_{\rm p} \geq 300$.

\subsection{Satellite galaxies}
\label{sec:satmodels}

For galaxies residing in the substructures of haloes we do not have information about the properties of their dark matter distribution. Therefore we have to rely ab initio on assumptions for both the shapes and orientations of satellite galaxies, which make up about $25\,\%$ of galaxies with $F814W<24$.

For early-type satellites we proceed in analogy to low-mass central galaxies and sample the axis ratios of three-dimensional ellipsoids from the histograms obtained for massive haloes with shape information at each redshift slice. Optionally these axis ratios are rescaled to mimic the use of the reduced inertia tensor. The ellipsoids are then oriented to point their major axis towards the central galaxy of the halo and subsequently projected along the line of sight using Eqs.$\,$(\ref{eq:ellipseprojection1}) to (\ref{eq:ellipseprojection3}) to yield image polarisations.

As an alternative model we implement the modifications of shapes and orientations of subhaloes found by \citet{knebe08} in high-resolution dark matter-only simulations (Knebe08 model hereafter). Their measures of triaxiality and sphericity of the satellite population can be converted to average axis ratios $\ba{c/a}_{\rm sat} = 0.80$ and $\ba{b/a}_{\rm sat} = 0.90$, which is in agreement with earlier works. \citet{knebe08} calculate the corresponding mean quantities for central galaxies from previous publications, obtaining $\ba{c/a}_{\rm cen} = 0.66$ and $\ba{b/a}_{\rm cen} = 0.76$, i.e. the satellite galaxies have more spherical shapes than central galaxies \citep[see also][]{kuhlen07}. 

We account for this by rescaling all axis ratios $\br{c/a}_{\rm cen}$ that are sampled from the histograms of central halo shapes via
\eq{
\label{eq:knebeshape}
\br{c/a}_{\rm sat} = 1 - \frac{1-\ba{c/a}_{\rm sat}}{1-\ba{c/a}_{\rm cen}}\; \bb{1-\br{c/a}_{\rm cen}}\;,
}
and likewise for $b/a$. Note that this formula is applied to results for the reduced inertia tensor as this was also used by \citet{knebe08}.

Late-type satellites are assumed to be thick circular disks (with their angular momentum perpendicular to the line connecting the position of the satellite with the centre of the halo) with the same properties as central disk galaxies, i.e. we create two models with $r_{\rm edge-on}=0.1$ and $r_{\rm edge-on}=0.25$, respectively.

\section{COSMOS galaxy shapes}
\label{sec:cosmos}

\subsection{Data}
\label{sec:cosmosdata}

We base our analysis on the \textit{HST} COSMOS Survey \citep{scoville07} which is the largest space-based survey to date and comes with excellent photometric redshift information (COSMOS-30, \citealp{ilbert08}). To estimate the intrinsic polarisation dispersion of galaxies as a function of brightness, galaxy type, and redshift, we make use of the shape catalogue produced for the weak lensing analysis by \citet{schrabback09}, matched with the ZEST morphological galaxy type classification\footnote{\texttt{http://irsa.ipac.caltech.edu/data/COSMOS/tables/morphology/}} \citep{scarlata07}.

We use only those galaxies that belong to the COSMOS-30 subsample, for which the absolute V-band magnitude and the photometric redshift are available. To approximately follow the magnitude limit of the ZEST sample ($I_{\rm AB}=24$), and to allow for an accurate treatment of noise (see below), we also impose a magnitude cut $F814W < 24$, which provides us with a total number of $88,600$ galaxies.

The ZEST classification assigns to galaxies type, bulgeness, irregularity, and other parameters using principal component analysis of morphological measures (see \citealp{scarlata07} for details). We define the following galaxy samples: early-type galaxies with TYPE$\;=1$ (elliptical) and IRRE$\;=0$ (regular morphologies only); bulge-dominated late-type galaxies with TYPE$\;=2$ (disk galaxies) and BULG$\;=0,1$, which includes most lenticular galaxies; disk-dominated late-type galaxies with TYPE$\;=2$ and BULG$\;=2,3$; irregular galaxies with TYPE$\;=3$.

For the galaxies in the matched catalogue the characteristic size in terms of the half-light radius, the observed magnitude in the $F814W$ band including an error estimate, and a LRG (luminous red galaxy) flag are also given. The latter identifies galaxies with $M_V<-19$ and a photometric type classifying them as \lq elliptical\rq\ (including S0) as the LRG sample of \citet{schrabback09}. This sample contains galaxies significantly fainter than $L^*$ and therefore not only LRGs in their standard definition. It is intended to comprise all galaxies with a potentially strong intrinsic alignment signal that could jeopardise cosmological analysis. We term this sample S10 LRGs henceforth and retain it as a complement to the morphologically defined galaxy samples. Note that some overlap with the ZEST early-type sample is expected.

\begin{figure}
\centering
\includegraphics[scale=.29,angle=270]{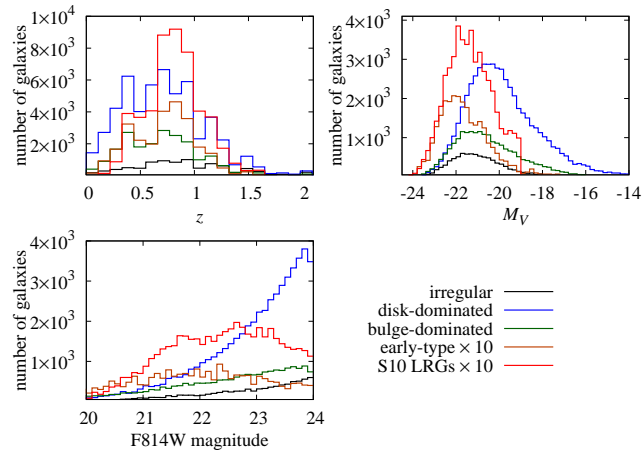}
\caption{Histograms for the COSMOS galaxy samples analysed in this work. S10 LRGs (see text for details) are shown as red lines (multiplied by a factor of 10 for easier inspection), early type galaxies as brown lines (also multiplied by a factor of 10), bulge-(disk-)dominated late-type galaxies as green (blue) lines, and irregular galaxies as black lines. Shown are the distributions with respect to $F814W$ apparent magnitude (bottom left), redshift $z$ (top left), and rest-frame magnitude $M_V$ (top right).}
\label{fig:distributions}
\end{figure} 

The galaxy ellipticities were determined with the KSB+ pipeline \citep{erben01}, which measures weighted second brightness moments of galaxy images. Great care has been taken to remove the effects of PSF smearing as well as spurious ellipticities due to image distortions, including spatial and temporal variations thereof. A detailed account of the shape measurement and tests of various systematics can be found in \citet{schrabback09}. 

The KSB scheme is designed to provide accurate and high signal-to-noise estimates of the gravitational shear rather than measure intrinsic galaxy ellipticities, giving a strong weight to the inner parts of the galaxy image. Note again that real galaxy images generally do not have a single ellipticity, but both absolute value and position angle of the ellipticity can be a function of radius. However, we will work with the single weighted polarisation estimate, once appropriately corrected, that is provided in the COSMOS shear catalogues. This needs to be kept in mind when confronting the COSMOS measurements with our simple, single-ellipticity galaxy shape models.

The KSB method provides PSF-corrected estimates of the galaxy ellipticity $\epsilon$, which is readily converted to polarisation via\footnote{Note that this relation holds for the ellipticities defined with \textit{unweighted} (or appropriately corrected) brightness moments.} $e = 2 \epsilon/(1+ |\epsilon|^2)$ \citep{bartelmann01}. While both definitions of galaxy ellipticity are equivalent, we choose $e$ in this paper as the measurement noise correction (see below) becomes slightly more convenient computationally, and as the ellipticity distributions display their features more clearly in terms of $e$.

The COSMOS galaxy shape catalogue is incomplete for very extended, and hence bright, objects because galaxies which do not fit well into the postage stamps used for shape measurement are discarded. Galaxy images with a half-light radius of $r_h=0.75^{\prime\prime}$ or larger were excluded before the KSB analysis, and a substantial fraction of objects with slightly smaller $r_h$ were subsequently flagged as having problematic shape measurements. As these objects have large angular size and high flux, PSF and noise effects are likely to be negligible.

Therefore we estimate the polarisation for these objects from axis ratios extracted from a \texttt{SExtractor} \citep{bertin96} catalogue with detection parameters optimised to include large bright galaxies, ignoring the PSF and setting the noise estimates for these measurements to zero. $24\,\%$ of all galaxies in the final matched catalogue are treated in this way. We find that including galaxies with \texttt{SExtractor} measurements does not alter the polarisation distributions in a statistically significant way.

The photometric redshift quality for the COSMOS-30 sample is excellent with $\sigma_z \approx 0.01(1+z)$ for galaxies with Subaru $i^+<24$ and $z<1.25$, degrading to $\sigma_z \approx 0.06(1+z)$ for the fainter galaxies at the maximum redshifts in our analysis around $z = 2$ \citep{ilbert08}. Photometric redshift scatter leads to the smoothing of any features in the signal when considered as a function of redshift or rest-frame magnitude (if the photometric redshift estimate is used to compute the distance modulus). However, as will be shown below, the modelled signals do not show strong features and are smoother than the observed ones even without including the effect of photometric redshift scatter, so that its influence can safely be neglected.

\subsection{Method}
\label{sec:cosmosmethod}

The polarisation dispersion is computed as
\eq{
\label{eq:sigmaeps}
\sigma_e = \sqrt{\ba{ e\, e^*}} = \sqrt{\frac{1}{N} \sum_{i=1}^N |e_i|^2}\;,
}
where $N$ is the number of galaxies in a given bin. We choose this quantity as our default measure because it provides information about the distribution of $e$ in a compact way and, besides, is of relevance to the noise computation in weak lensing two-point statistics \citep{kaiser92}.

The polarisation dispersion measured from COSMOS data is composed of the intrinsic polarisation dispersion and contributions from measurement noise. To correct for the latter, we use a modified version of the Fisher matrix approach proposed by \citet{leauthaud07}. Assuming that the light distribution in the image can be described by a bivariate Gaussian whose covariance is given by the second-order brightness tensor, one obtains an estimate of the error on the brightness moments which can be propagated into an error on $\sigma_e$. Since the Fisher matrix provides us with an expectation value, we do not need to revert to the actual galaxy images but require only the brightness moments, the apparent magnitude, and the S/N for each galaxy as input. Full details of this procedure are provided in Appendix \ref{app:noise}.

\begin{figure}
\centering
\includegraphics[scale=.34,angle=270]{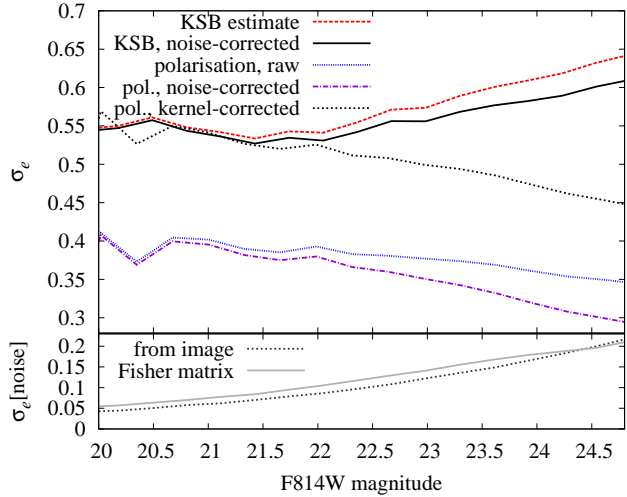}
\caption{Polarisation dispersion $\sigma_e$ as a function of apparent $F814W$ magnitude. \textit{Top panel}: The blue dotted line corresponds to $\sigma_e$ measured from the \lq raw\rq\ polarisations in the shear catalogue, the red dashed line to $\sigma_e$ obtained from the KSB shear estimates, and the black solid line to the noise-corrected $\sigma_e$. We also process the raw polarisations directly, resulting in a noise-corrected $\sigma_e$ shown as violet dot-dashed line, and $\sigma_e$ that has subsequently been corrected for the circularisation due to the Gaussian weight in the brightness moments, see the black dotted line. Note the good agreement between the two black curves at bright magnitudes. \textit{Bottom panel}: Dispersion $\sigma_e$ of the measurement noise, as estimated from the images according to \citet{hoekstra00}, see the black dotted line, and from our Fisher matrix formalism, see the grey solid line. The two approaches show very good agreement.}
\label{fig:epserror}
\end{figure} 

In a subset of the COSMOS field we also calculate the measurement noise directly from the background noise of individual galaxy images, using the method outlined in the appendix of \citet{hoekstra00}. The resulting measurement noise contribution to $\sigma_e$ at different $F814W$ magnitudes is shown in the bottom panel of Fig.$\,$\ref{fig:epserror}, together with the prediction of the Fisher matrix formalism. The agreement of the two approaches is very good down to our magnitude limit, which is remarkable given that our version of the Fisher matrix calculations does not have any free parameters, as opposed to the \citet{leauthaud07} formalism.

In Fig.$\,$\ref{fig:epserror}, top panel, we have plotted $\sigma_e$ as determined from the output of the KSB shear estimation pipeline, as well as after subtracting in quadrature the measurement noise from the Fisher estimates. As expected, the difference between the two curves is close to zero at bright magnitudes where galaxies have high signal-to-noise wherefore measurement noise is negligible. This difference gradually increases for fainter magnitudes until the dispersion of the measurement noise reaches about 0.2, which marks a $10\,\%$ contribution to $\sigma_e^2$ at $F814W=25$.

We compare this result to Fig.$\,$17 of \citet{leauthaud07} who used galaxy shape measurements of COSMOS galaxies based on the method proposed by \citet{rhodes00}. Note however that the authors plotted the mean dispersion of the ellipticity components $\epsilon_{1,2}$ rather than the dispersion of the total polarisation $e$ as done in this work. Generally, we find good agreement but obtain overall slightly lower values of the noise-corrected ellipticity dispersion, despite smaller noise-corrections (for $F814W<21.5$ we find a shear dispersion per component of 0.23 compared to $\sim 0.24$ in \citealp{leauthaud07}). 

Small differences in the ellipticity dispersion are not unexpected due to the different shape measurement methods used. For instance, in the KSB implementation of \citet{schrabback09} the shear tensor is individually determined for each galaxy instead of for ensembles of galaxies as in \citet{leauthaud07}. Very similar to our Fig.$\,$\ref{fig:epserror}, \citet{leauthaud07} also find a slight increase of $\sigma_e$ as a function of apparent magnitude, even after noise correction, which, if physical, could be caused by changes in the galaxy population (e.g. due to redshift evolution or changes in the fraction of early- and late-type galaxies) at these magnitudes.

To cross-check the polarisation measurement by means of KSB shear estimates, we also determine $\sigma_e$ via an alternative route: We start from the \lq raw\rq\ polarisations directly obtained from the observed brightness moments and correct them for measurement noise via the methods outlined above. The resulting polarisations are still affected by PSF smearing as well as spurious ellipticity introduced by image distortions, but for bright and extended objects whose apparent size is large compared to the PSF full width at half maximum the polarisation should be similar to the PSF-corrected one.

However, the brightness moments are computed with a circular Gaussian kernel \citep[see][]{schrabback09}, with the consequence that, even for bright galaxies, the measured absolute value of the polarisation $|e|$ will be significantly smaller than the \lq true\rq\ value, as e.g. measured from isophotes (which is impossible to determine with sufficient accuracy for the majority of faint and small galaxies in COSMOS). In Appendix \ref{app:ksbkernel} we show how to correct for the circularisation of the brightness moments by quantifying the effect using analytic Sersic light profiles\footnote{Note that in the KSB scheme this step corresponds to the division by the \lq shear tensor\rq\ which accounts for the circularisation due to the Gaussian kernel as well as due to the PSF \citep[e.g.][]{bartelmann01}.}.

As one can see in the top panel of Fig.$\,$\ref{fig:epserror}, the raw and noise-corrected polarisations yield very small $\sigma_e < 0.4$, which is caused by the circularisation effects of the PSF and the circular Gaussian weight in the brightness moments. Correcting for the latter indeed increases $\sigma_e$ to be in very good agreement with the result of our default approach. Beyond $F814W \approx 22$ however, the curves diverge quickly as the PSF smearing for these fainter and smaller galaxy images becomes an increasingly important effect.

It is also instructive to compare our results for the raw polarisation with those of Fig.$\,$20 in \citet{hoekstra00} who found that the dispersion is constant with apparent magnitude, which they traced back to the fact that increasing PSF effects at fainter magnitudes happen to be exactly balanced by an increase in measurement noise. Their measurements were based on shallower \textit{HST} imaging, so that for COSMOS we expect the PSF effects to win over noise for $F814W > 22$ where the noise estimates by \citet{hoekstra00} become substantial. This is indeed the case as evident by the downturn of $\sigma_e$ at $F814W \approx 22$.

Once we have determined the final, noise-corrected polarisations for all galaxies, the polarisation dispersion is calculated as a function of rest-frame magnitude $M_V$ and redshift $z$. Additionally, we divide the galaxies up according to the various morphological and photometric type classifications discussed in Sect.$\,$\ref{sec:cosmosdata}.

The error on the dispersion is estimated by bootstrapping from 50 catalogues containing the same number of galaxies as the input catalogue (after the selection criteria in magnitude and morphology have been applied). The corresponding statistics from the simulations are constructed as follows. Galaxies are randomly resampled from the simulation catalogues according to histograms in the $(M_V,z)$ grid for the different samples. While $M_V$ rest-frame magnitudes are directly available from the semi-analytic models, we reproduce the type classification via the $R_{\rm type}$ parameter. As the S10 LRGs sample encompasses elliptical and lenticular galaxies, we set $R_{\rm type}>0.4$ while for the early-type sample $R_{\rm type}>0.6$ is used. The ZEST Type$=2$ class includes most of the S0 galaxies, so that for our bulge-dominated sample we set $R_{\rm type}<0.6$. Finally, $R_{\rm type}<0.4$ is chosen for the disk-dominated sample. Note that we do not attempt to model irregular galaxies.

We build 5 catalogues for each of the 64 lines of sight with numbers of galaxies chosen such that the total in each line of sight is of the same order as the number of galaxies in the COSMOS catalogues. This ensures that the statistical constraints from the simulations are at least as good as from the observations, and that repeated draws of galaxies from regions in the $(M_V,z)$ grid with sparse sampling are kept to a minimum. We show and analyse only bins which contain 30 galaxies or more after applying all cuts.

In addition to the selection in $M_V$ and $z$ we have to account for the apparent magnitude cut in $F814W$. As this filter is not available in our simulation catalogues, we approximate it by the SDSS bands $i$ and $z$ which roughly cover the same wavelengths. Resorting to the tables of \citet{fukugita96}, we indeed find that the colour $(i+z)/2-F814W$ evolves little with redshift. It is essentially constant at 0.31; only for early-type galaxies above $z=0.5$ does this value start to decrease moderately. With this conversion we impose the cut $F814W<24$ on our models.

\section{Results}
\label{sec:cosmosresults}

\subsection{Polarisation dispersion}
\label{sec:poldispersion}

\begin{figure*}
\centering
\includegraphics[scale=.34,angle=270]{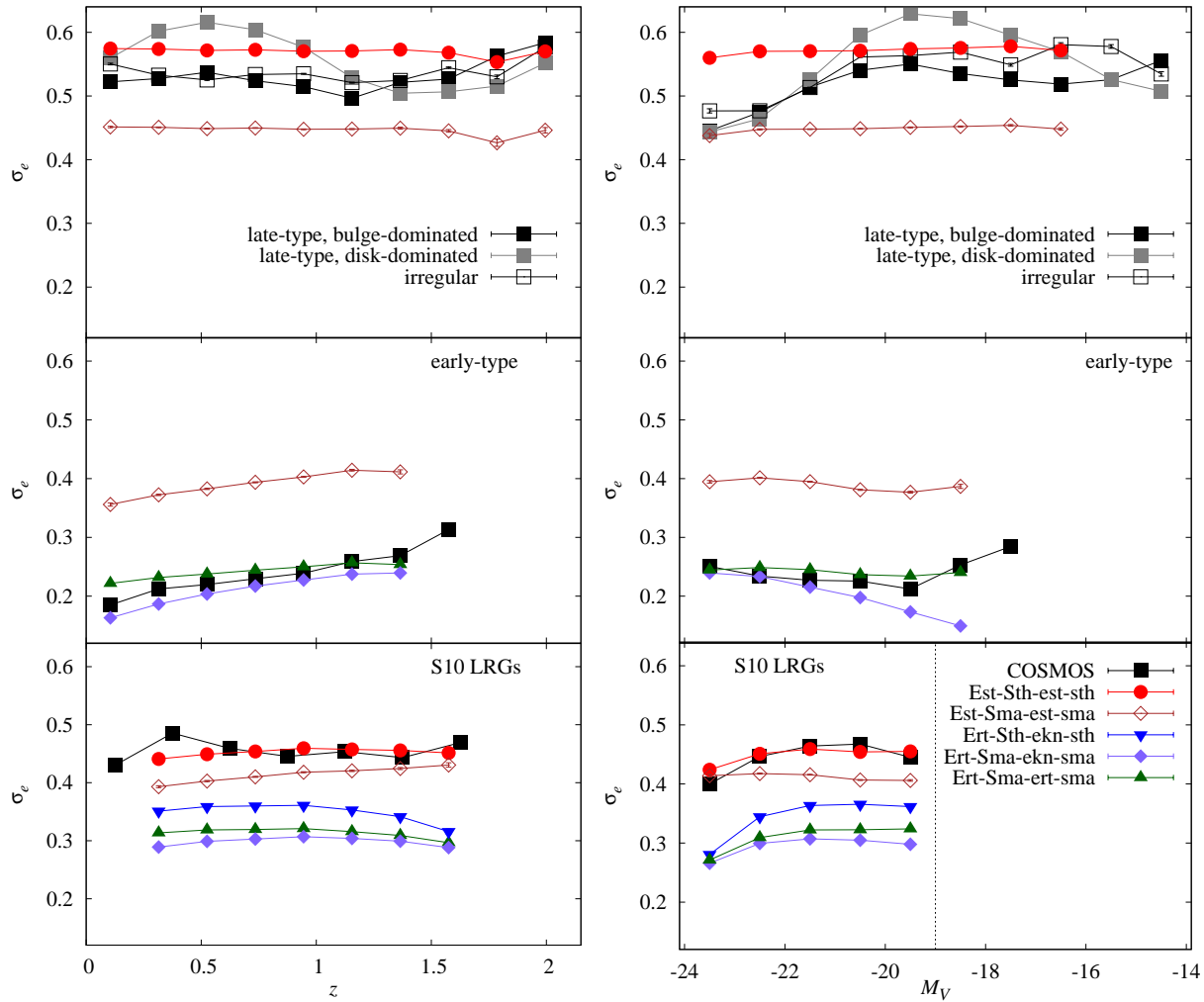}
\caption{Polarisation dispersion $\sigma_e$ as a function of redshift (left panels) and rest-frame $V$-band magnitude (right panels). The top panels display results for the late-type sample, the centre panels for the early-type sample, and the bottom panels for the S10 LRG sample. The dispersion based on the noise-corrected polarisations from COSMOS are shown as black filled squares (grey filled squares for disk-dominated late types; black open squares for irregular galaxies) for an apparent magnitude cut at $F814W=24$. The corresponding results for the simulation-based models are shown as red circles (brown diamonds) when basing early-type models on the simple inertia tensor [\texttt{Est;est}] and using late-type galaxies with $r_{\rm edge-on}=0.1$ [\texttt{Sth;sth}] ($r_{\rm edge-on}=0.25$ [\texttt{Sma;sma}]). The same late-type models combined with elliptical galaxies based on reduced inertia tensor measurements including the Knebe08 model for satellites [\texttt{Ert;ekn}] are given by the blue downward triangles ($r_{\rm edge-on}=0.1$) and the violet diamonds ($r_{\rm edge-on}=0.25$), respectively. Discarding the modifications due to the Knebe08 model changes the model from the violet diamonds to the green upward triangles. Note that the S10 LRG sample has been defined with a magnitude cut $M_V<-19$, as indicated by the black dotted line. Error bars given by the mean field-to-field variation are shown throughout, but remain much smaller than the size of the symbols. The simulation results for the disk- and bulge-dominated late-type sample are very similar, so that we only show the latter in the top panel.}
\label{fig:cosmos}
\end{figure*} 

The resulting polarisation dispersions $\sigma_e$ for the COSMOS data as well as for various galaxy shape models based on the Millennium Simulation are shown in Fig.$\,$\ref{fig:cosmos}, as a function of $V$-band rest-frame magnitude and redshift. We distinguish between models in which the shapes of early-type central and satellite galaxies are either computed from the simple [\texttt{Est;est}] or the reduced [\texttt{Ert;ert}] inertia tensor (see Table \ref{tab:galaxymodels} for a list of the model identifiers). Moreover the Knebe08 model for early-type satellites is used in combination with reduced inertia tensor shapes for central early types [\texttt{Ert;ekn}]. For central and satellite late-type galaxies we vary the edge-on axis ratio between $r_{\rm edge-on}=0.1$ [\texttt{Sth;sth}] and $r_{\rm edge-on}=0.25$ [\texttt{Sma;sma}].

We observe clearly distinct ranges of $\sigma_e$ for the COSMOS late, early, and S10 LRG samples with $\sigma_e \ga 0.5$, $\sigma_e < 0.3$, and $0.4<\sigma_e <0.5$, respectively. The dependence on redshift and luminosity is generally weak and not monotonic in most cases. The model \texttt{Est-Sth-est-sth} (red circles) reproduces very well the polarisation dispersion of the S10 LRG sample and predicts the correct order of $\sigma_e$ for the late-type samples, without showing any of their variation in $z$ and $M_V$. The \texttt{Est}-based models largely overpredict $\sigma_e$ for the early-type sample. Instead the models based on rounder halo shapes (\texttt{Ert,ert/ekn}), which generally yield lower dispersions, reproduce the COSMOS observations well. Note that incorporating the additional rounding of early-type satellites as suggested by the Knebe08 model only marginally lowers $\sigma_e$ at low redshift and fainter magnitudes where a larger fraction of satellite galaxies contributes to the signal.

\begin{figure}
\centering
\includegraphics[scale=.335,angle=270]{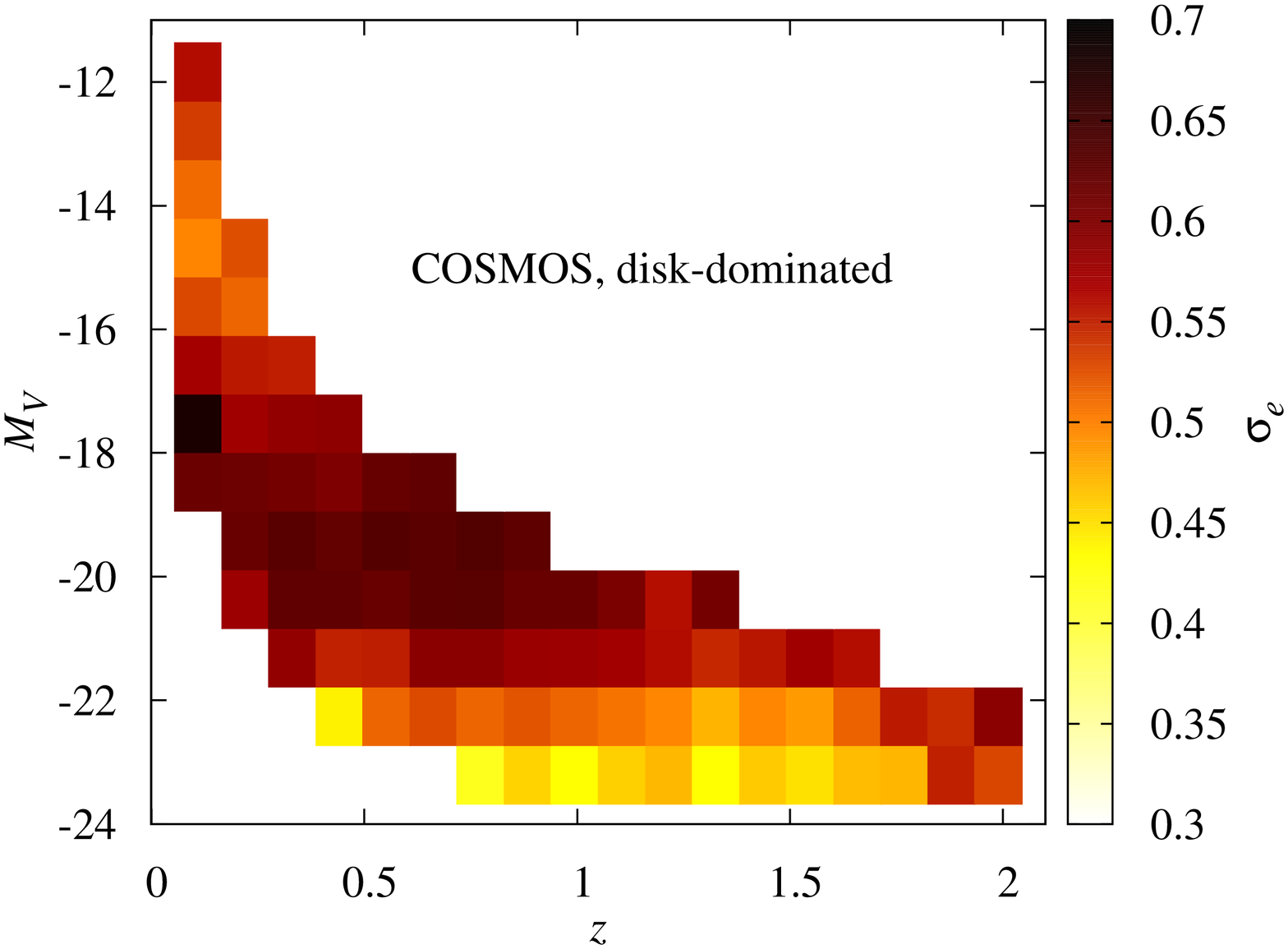}
\caption{Polarisation dispersion $\sigma_e$ as a function of rest-frame $V$-band magnitude and redshift for the COSMOS disk-dominated late-type sample with $F814W < 24$. There is a tendency towards lower $\sigma_e$ for luminous galaxies with $M_V \la -22$ in a broad redshift range $0.5 \la z \la 1.8$, and towards higher $\sigma_e$ for galaxies in the range $-21 \la M_V \la -18$. The former trend dominates at low redshift, the latter at high redshift.}
\label{fig:shapevar_2D}
\end{figure}

The three late-type samples shown in the top panel feature similar signals, in particular they share a pronounced decrease in $\sigma_e$ for $M_V \la -21$, which is not seen in any of the simulation-based models. The observations also display small variations in the redshift dependence, strongest in the disk-dominated sample with a shallow maximum at $z \sim 0.5$ and a minimum around $z \approx 1.5$. Similar small-amplitude variations of the redshift dependence were observed by \citet{leauthaud07}. To investigate these discrepancies, we have plotted the two-dimensional distribution of $\sigma_e$ for the disk-dominated sample in Fig.$\,$\ref{fig:shapevar_2D}.

The polarisation dispersion strongly decreases down to $\sigma_e \approx 0.4$ for $M_V \la -21$ over almost the complete redshift range that the observations cover, i.e. for $0.5 \la z \la 1.8$. Conversely, there is a region of high $\sigma_e \sim 0.6$ in the range $-21 < M_V < -17$ that dominates redshifts below unity. This explains the sinusoidal $z$ dependence observed for this sample in Fig.$\,$\ref{fig:cosmos}, upper left panel. We obtained similar results for photometrically selected galaxy samples based on \citet{mobasher07}.

As the decrease of $\sigma_e$ for bright objects is seen across all late-type samples, as well as for morphological and photometric type selections, it seems unlikely that this could be caused by contamination by earlier galaxy types. However, we note that brightest cluster galaxies which have recently formed stars (as e.g. studied by \citealp{bildfell08}) may consistently be classified as late types but would actually reside in early-type haloes; see the $\sigma_e$ values of the \texttt{Est-Sth-est-sth} model for the S10 LRG sample.

Generally, it seems reasonable that very bright disk galaxies, which preferentially reside in high-density regions, have distinctively different morphological properties compared to field galaxies. It is interesting to note that $\sigma_e$ for $M_V<-22$ reaches values very close to those predicted by the models that include disk galaxies with $r_{\rm edge-on}=0.25$, which was motivated by the existence of bulges. Hence the observed decrease in $\sigma_e$ with increasing luminosity could actually be a transition to intrinsically \lq thicker\rq\ galaxies, although it remains unclear why all late-type samples are affected in the same way.

\subsection{Polarisation distributions}
\label{sec:poldistribution}

While the variance of the complex polarisation is a convenient variable to assess the statistical properties of $e$, one can make an attempt at comparing the full distributions obtained from models and observations directly. Since the correction for measurement noise outlined in Section \ref{sec:cosmosmethod} is statistical in nature, we cannot de-noise the polarisation measurements of individual galaxies. Thus we compare the distributions of $|e|$ as produced by the shear pipeline, i.e. including all corrections for PSF effects, but also measurement errors due to noise. To allow for a fair comparison, we devise a simple noise model and apply it to the model polarisation distributions as follows.

We divide the different COSMOS galaxy type samples into ten equidistant bins in $|e|$ and create for each bin a histogram of the measurement error on $|e|$ computed from our Fisher matrix formalism. Each polarisation from the simulation-based catalogues is then modified by a shift randomly drawn from a zero-mean Gaussian that has a width equal to the measurement error which is in turn randomly sampled from the corresponding histogram. Resulting polarisations with $|e|<0$ or $|e|>1$ are discarded and resampled, so that, effectively, the scatter due to measurement noise preferentially increases the ellipticity of nearly round objects and decreases the ellipticity of galaxies viewed nearly edge-on, as one would expect in reality.

\begin{figure}
\centering
\includegraphics[scale=.35,angle=270]{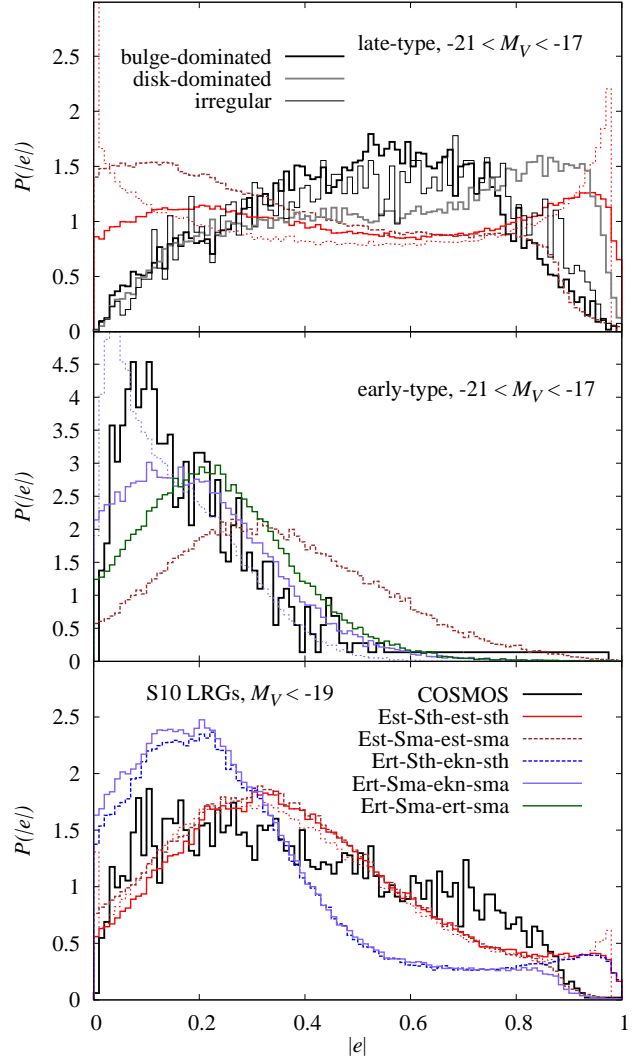}
\caption{Distribution of the absolute value of the polarisation $|e|$. \textit{Top panel}: Probability density of $|e|$ for the COSMOS late-type samples in the rest-frame magnitude range $-21 < M_V < -17$. The curve for the polarisation as output by the KSB pipeline (i.e. without measurement noise corrections) is shown as thick black (thick grey; thin black) line for the bulge-dominated late types (disk-dominated late types; irregular galaxies). The solid and dashed curves resulting from the simulation-based models with a model for measurement noise added are assigned the same colour coding as used in Fig.$\,$\ref{fig:cosmos}. The dotted red curve represents the result for the \texttt{Est-Sth-est-sth} model without noise. \textit{Centre panel}: Same as above, but for the early-type sample. The dotted violet curve represents the result for the \texttt{Ert-Sma-ekn-sma} model without noise. \textit{Bottom panel}: Same as above, but for the full S10 LRG sample. Again, the dotted red curve represents the result for the \texttt{Est-Sth-est-sth} model without noise.}
\label{fig:cosmos_histo}
\end{figure} 

In Fig.$\,$\ref{fig:cosmos_histo} we have plotted the probability distributions of $|e|$ for the COSMOS late, early, and S10 LRG samples, limiting the magnitude range to $-21<M_V<-17$ in all but the last case. For the S10 LRG sample the models that were close to the COSMOS results in terms of $\sigma_e$ (see the bottom panels of Fig.$\,$\ref{fig:cosmos}) also perform well in reproducing the full distribution of $|e|$. The \texttt{Est-Sth-est-sth} model distribution is slightly more compact, compensated by a small excess of galaxies at $|e|>0.9$. Most of these highly elliptical objects do not yield a reliable shape measurement and are hence absent in the observational distributions. The models that include rounder early-type galaxy shapes based on reduced inertia tensor measurement for the dark matter haloes are inconsistent with the data at high significance, irrespective of what is assumed for disk galaxies.

{The impact of the noise model on the probability distributions of $|e|$ is illustrated by comparing the two red curves of the \texttt{Est-Sth-est-sth} model in the bottom panel of Fig.$\,$\ref{fig:cosmos_histo}, where the dotted one is without noise. The scatter due to noise smooths sharp features in the distributions, such as the small peak due to disk galaxies viewed edge-on at $|e| \la 1$. Moreover polarisation values are re-distributed away from extreme values close to 0 and 1 although this effect is small for the bright S10 LRG sample due to low measurement noise.

The polarisation distribution of the early-type sample is strikingly different, with hardly any values above $|e|=0.6$ and a strong peak around $|e|=0.1$, resulting in the very small dispersion observed in Fig.$\,$\ref{fig:cosmos}. Again, the models which fared well in reproducing the redshift and luminosity dependence of $\sigma_e$ also yield distributions that are close to the observations, particularly the combination \texttt{Ert,ekn}. However, the comparison with the modelled distribution without noise, which is more strongly peaked, suggests that we might slightly over-estimate measurement noise for this sample.

The late-type samples feature quite broad distributions of $|e|$, explaining the high values of $\sigma_e$ seen in Fig.$\,$\ref{fig:cosmos}. All three samples share a pronounced deficit of nearly circular galaxy images, which is clearly discrepant with the model predictions\footnote{\refmark{Note that the distributions of late-type galaxies with $M_V<-21$ also show this deficit. The decrease in $\sigma_e$ seen in Fig.$\,$\ref{fig:cosmos} for these objects is caused by a shift of the peak of the polarisation distributions from $|e|>0.5$ towards $|e|<0.5$.}}. At low $|e|$ the models are dominated by face-on disks; see the noise-free distribution given by the dotted red line in Fig.$\,$\ref{fig:cosmos_histo} which is in good agreement with the analytic prediction for randomly oriented thin disks given by Eq.$\,$5.20 of \citet{bernstein02}. Measurement noise is comparatively large for the late-type samples, and its impact is strongest at low $|e|$, which leads to a significant re-distribution of polarisations towards values much larger than zero due to the skewness of the noise distribution (compare the dotted and solid red curves).

This trend due to measurement noise suggests that a moderate under-estimation of noise for the late-type samples could partly remedy the discrepancy between model and observations. Such an under-estimation could e.g. be caused by our ad-hoc assumption of a truncated Gaussian for the probability distribution of noise per galaxy. However, if noise were the only reason for the observed discrpancy of polarisation distributions, one would expect good agreement between model and COSMOS data for a bright, and thus low-noise, subsample. For disk-dominated late-type galaxies with $F814W<22$ the fraction of polarisations with $|e|<0.2$ increases by only $30\,\%$ compared to the sample with a limiting magnitude of $F814W<24$, so that measurement noise does not fully explain the differences between model and observation.

\citet{bernstein02} presented similar intrinsic polarisation distributions for bright, low-noise galaxies in the CTIO lensing survey. Interestingly, their low surface brightness sample, dominated by spiral galaxies, also contains few near-circular galaxies, with a strong decline of the probability distribution of $|e|$ below $|e|=0.1$. The authors concluded that the disks of these late-type galaxies are not perfectly circular. Indeed, changing the axis ratio of the light distribution of a late-type galaxy viewed face-on from 1 to 0.9 already translates into a minimum ellipticity of $|e| \geq 0.1$. This could readily be achieved in practice by the presence of luminous substructure such as giant star-forming regions or blended satellite galaxies. Furthermore it should be kept in mind that KSB-like methods put a strong weight on the inner parts of a galaxy in measuring brightness moments, so that features like prominent bars may play a non-negligible role. 

However, since the same deficit of galaxies with low $|e|$ is observed across our three late-type samples, which vary considerably in the level of irregularity in the light distribution as well as the prominence of bulges, any of the aforementioned explanations in terms of galaxy morphology seem unlikely to be conclusive.

The image size cut in the weak lensing shape measurement procedure cannot explain this deficit either because adding the substantial fraction of \texttt{SExtractor} ellipticities (see Sect.$\,$\ref{sec:cosmosdata}) does not change the distributions qualitatively. However, both \texttt{SExtractor} and KSB shape measurements can fail if cosmic rays, bad pixels, diffraction spikes, and other artefacts affect the galaxy image, which is more likely to happen if the area covered by the image is large. For a given half-light radius, this effect therefore preferentially discards near-circular objects from further analysis and so could potentially contribute to the observed deficit.

Another potential issue that was not considered in our analysis is the pixelisation of the light distribution. With an average half-light radius of $7.8$ pixels for the disk-dominated late-type sample, a pixelated galaxy image is likely to have a small residual ellipticity even if the actual isophotes are perfectly circular. In our case the impact of pixelisation seems negligible though, as the early-type sample does not feature a deficit of low-ellipticity objects but has on average even smaller half-light radii ($6.1\,$px).

At high $|e|$ the model distributions are governed by edge-on galaxies and hence by the choice of $r_{\rm edge-on}$. Despite our simplistic choice of a single disk thickness for all galaxies, we find fair agreement between the disk-dominated COSMOS sample and the model with $r_{\rm edge-on}=0.1$ at least for the largest polarisations, and good agreement between the bulge-dominated COSMOS sample and the model with $r_{\rm edge-on}=0.25$ for $|e|>0.9$, which justifies these model assumptions.

\section{Conclusions}
\label{sec:conclusions}

In this work we analysed the statistical properties of galaxy ellipticities, confronting samples from the \textit{HST} COSMOS Survey with a suite of models based on the dark matter halo and galaxy properties provided by the Millennium Simulation. The galaxy shape models differentiate between central and satellite as well as early- and late-type galaxies, and incorporate additional information on the link between luminous and dark matter from other simulations. We confirm earlier work in that at least 300 particles per halo are required to measure accurate three-dimensional shapes (as well as angular momenta, see \citealp{bett07}), implying deviations in the Cartesian components of the projected ellipticity of at most 0.02.

Intrinsic galaxy ellipticities (measured in terms of weak lensing polarisation) for a sample of about $90,000$ COSMOS galaxies were extracted from gravitational shear estimates based on second brightness moments, taking full advantage of the scrutiny and systematics testing undertaken by \citet{schrabback09}, particularly with respect to the smearing of the PSF and spurious ellipticities introduced by telescope and camera. We demonstrated that the resulting polarisation dispersion $\sigma_e$ is robust with respect to the methods used to correct for the effects of the circular kernel in the brightness moments, as well as measurement noise. To estimate the latter, we devised a Fisher matrix formalism that yields accurate results without the need to resort to observed or simulated images.

Splitting the COSMOS galaxies into several samples according to type, we detect a significant dichotomy in the polarisation dispersion between early- and late-type galaxies, the latter having $\sigma_e$ of up to a factor of 2 larger. We find no evidence for a redshift evolution of $\sigma_e$ in any sample, which is in agreement with earlier work \citep{leauthaud07}. The dependence of $\sigma_e$ on luminosity is also weak, except for the brightest (rest-frame $M_V \la -21$) late-type galaxies which feature a decrease in $\sigma_e$ of 0.1 to 0.2 over a broad range in redshifts out to $z \approx 1.8$. This effect is present in all late-type samples considered and persists when selecting galaxies according to photometric rather than morphological properties. Future investigations that measure a dependence of $\sigma_e$ on local density will be able to establish whether this finding could hint at an environment dependence, with lower ellipticities found in high-density regions.

Studying the distributions of absolute values of the polarisation, we generally find fair agreement between observations and those models that also fit $\sigma_e$ well. A notable exception is the low fraction of close to circular galaxies ($|e| \la 0.2$) in the late-type samples, which was also observed by \citet{bernstein02}. The deficit of nearly circular galaxy images is partly caused by the strong upward scatter of $e$ due to measurement noise. We also identify selection effects implicated by image artefacts as a potentially important contribution. Both hypotheses can be verified in forthcoming work via forward modelling using realistic image simulations.

The considerable variation of $\sigma_e$ with galaxy properties by up to a factor of 2 in extreme cases (see Fig.$\,$\ref{fig:cosmos}) suggests that a better understanding of the physics that drive the observable shapes of galaxies might help optimising weak lensing survey designs. For instance, switching from one of the COSMOS late-type samples (which are similar to a typical weak lensing sample) to one similar to the early-type sample would yield the same shape noise level if the number density of the latter were about a factor of 4 smaller (ignoring the contribution by measurement noise which can be independently controlled, e.g. by apparent magnitude cuts). Of course, early-type galaxies form only a small percentage of the total number of galaxies suitable for weak lensing measurement, but they may have additional favourable properties, e.g. in terms of the quality of photometric redshifts attainable, the anticipated good constraints on intrinsic alignment contamination, or the presence of colour gradients across galaxy images.

Perhaps more importantly, detailed knowledge of the intrinsic ellipticity distribution of a weak lensing galaxy sample is essential to understand and control noise-induced biases in shape measurement, independently of the method applied \citep{melchior12}. A considerable variation of the intrinsic ellipticity distribution with galaxy type, as found in this work, implies that any calibration sample to obtain intrinsic ellipticities must be carefully chosen to be representative of the full sample.

In spite of the necessarily still simplistic assumptions about the shapes of galaxies and how these are linked to the dark matter properties, we could always identify a simulation-based model that was capable of yielding good agreement with the observed $\sigma_e$. The difference in the amplitude of $\sigma_e$ between the S10 LRG and late-type samples is reproduced quantitatively, including the independence on redshift, by the \texttt{Est-Sth-est-sth} model which assumes disks with a thickness-to-diameter ratio $r_{\rm edge-on}=0.1$ and elliptical galaxy shapes based on the simple inertia tensor of haloes.

Which model describes the observations best seems to depend strongly on the sample selection. The morphologically selected sample of regular ellipticals is well represented by the more spherical shapes determined from reduced inertia tensor measurements and the Knebe08 prescription for satellites while the photometrically selected S10 LRG sample of early types including lenticular galaxies (in both data and models) is accurately fit by models with shapes based on the simple inertia tensor. In the context of weak lensing surveys galaxy type selection via photometry is likely to be more relevant as these come automatically with the measurement of photometric redshifts.

We included two values of $r_{\rm edge-on}$ into our galaxy disk models, a low value motivated by pure disks and a higher value that qualitatively takes into account the effect of a bulge viewed edge-on. The comparison of the distributions of high values of the polarisation between models and COSMOS late-type data suggests that these values provide a fair representation of disk- and bulge-dominated spiral galaxy populations. A more realistic model for late-type galaxies could sample from a distribution of values of $r_{\rm edge-on}$, obtained from a representative sample in the local Universe.

Incorporating the shape properties of disturbed and irregular galaxies is relevant as they constitute an increasingly important fraction of the full (weak lensing) sample at redshifts around unity and above \citep{abraham96,bundy05}. While it remains unclear how to devise an explicit shape model for these galaxies, the observed properties of polarisation dispersion and distributions of the irregular sample match closely those of the other late-type samples, so that for practical purposes they can be treated as part of a population modelled as disk galaxies.

The approach taken in this paper can readily be applied to higher resolution simulations, where shapes and angular momenta of individual haloes are obtained also for satellite galaxies. Further improvements are expected from more observations of ensemble properties of galaxy shapes (such as for $r_{\rm edge-on}$, see above), as well as from more and larger hydrodynamic simulations which give insight into the connection between the shapes of the dark matter halo and the luminous galaxy. 

But even at the present stage, the analysis of one-point statistics of galaxy shapes in the form of the polarisation dispersion has proven to be insightful and capable of discriminating between models with high significance. Compared to measurements of correlations of intrinsic galaxy shapes, dispersions and distributions of $e$ can be obtained deeply into high-density (non-linear) regions, are not affected by galaxy bias, and are possibly less susceptible to sample variance. Hence, to constrain models of intrinsic galaxy shapes and alignments, the approach presented here is a valuable complement to second-order statistics of galaxy shapes.

\section*{Acknowledgments}

We thank Catherine Heymans and Andy Taylor for helpful discussions, and our referee for useful suggestions that helped improving the paper.

BJ acknowledges support by STFC and a UK Space Agency Euclid grant. ES and HH acknowledge the support of the Netherlands Organization for Scientific Research (NWO) grant number 639.042.814 and support from the European Research Council under FP7 grant number 279396. PEB acknowledges support by the Deutsche Forschungsgemeinschaft (DFG) under the project SCHN 342/7–1 in the framework of the Priority Programme SPP-1177, and the Initiative and Networking Fund of the Helmholtz Association, contract HA-101 (‘Physics at the Terascale’). JH, SH, and PS acknowledge support by the DFG through the Priority Programme 1177 `Galaxy Evolution' (SCHN 342/6 and WH 6/3) and the Transregional Collaborative Research Centre TRR 33 `The Dark Universe'. SH also acknowledges support by the National Science Foundation (NSF) grant number AST-0807458-002. TS acknowledges support from NSF through grant AST-0444059-001, and the Smithsonian Astrophysics Observatory through grant GO0-11147A.

The simulations used in this paper were carried out as part of the programme of the Virgo Consortium on the Regatta supercomputer of the Computing Centre of the Max-Planck-Society in Garching, and the Cosmology Machine supercomputer at the Institute for Computational Cosmology, Durham. The Cosmology Machine is part of the DiRAC Facility jointly funded by STFC, the Large Facilities Capital Fund of BIS, and Durham University.

This research has made use of the NASA/IPAC Infrared Science Archive, which is operated by the Jet Propulsion Laboratory, California Institute of Technology, under contract with the National Aeronautics and Space Administration.

\appendix

\section{Parallel projection of an ellipsoidal halo}
\label{app:ellipse}

We define the shape of a dark matter halo and the corresponding early-type galaxy in terms of the eigenvectors and eigenvalues of the halo inertia tensor $\mathbf{M}$, given by Equation (\ref{eq:massquadrupole}). Therefore the surface of the ellipsoid that serves as the model of the galaxy is constituted by all points with coordinates $\vek{x}$ which fulfil
\eq{
\label{eq:ellispoiddef}
\vek{x}^\tau\; \mathbf{M}^{-1}\; \vek{x} = 1\;.
}
The line of sight is chosen to coincide with the third coordinate axis along which we parallel project the ellipsoid. To this end we define sets of straight lines
\eq{
\label{eq:rays}
\vek{x}(t) = \vek{y} + t\, \vek{n}\; ~~\mbox{with}~~ \vek{n} = \bc{0,0,1}^\tau; ~ \vek{y} = \bc{\vek{y}_\perp^\tau,0}^\tau,
}
i.e. these lines are parallel to the line of sight and, for $t=0$, intercept the plane onto which the ellipsoid is projected at position $\vek{y}_\perp$.

Inserting Equation (\ref{eq:rays}) into Equation (\ref{eq:ellispoiddef}), one obtains a quadratic equation in $t$ of the form
\eqa{
\label{eq:quadraticeq}
A\,t^2 + B\,t + C \!\!&=\!\!& 0 ~~\mbox{with}~\\ \nn
&& \hspace*{-2.5cm} A=\vek{n}^\tau\, \mathbf{M}^{-1}\, \vek{n}\;; ~B=2\, \vek{n}^\tau\, \mathbf{M}^{-1}\, \vek{y}\;; ~C=\vek{y}^\tau\, \mathbf{M}^{-1}\, \vek{y} - 1\;.
}
Those lines that touch the surface of the ellipsoid in one point are selected by requiring that the discriminant vanishes, $4AC=B^2$. This condition is equivalent to
\eqa{
\label{eq:projectedellipse}
1 &=& \vek{y}^\tau\, \mathbf{M}^{-1}\, \vek{y} - \frac{\br{\vek{n}^\tau\, \mathbf{M}^{-1}\, \vek{y}}^2}{\vek{n}^\tau\, \mathbf{M}^{-1}\, \vek{n}}\\ \nn
&=& \vek{y}^\tau \bb{ \mathbf{M}^{-1} - \frac{\mathbf{M}^{-1}  \vek{n}  \vek{n}^\tau \mathbf{M}^{-1} }{\vek{n}^\tau\, \mathbf{M}^{-1}\, \vek{n}} } \vek{y}\\ \nn
&\equiv& \vek{y}_\perp^\tau\; {\mathbf W}^{-1}\; \vek{y}_\perp\;.
}
In the last line we have defined the symmetric two-dimensional tensor ${\mathbf W}$. This expression is the defining equation of an ellipse that lies in the plane of the sky.

Note that our procedure is not equivalent to simply projecting the inertia tensor ${\mathbf M}$ along the line of sight as this would only correspond to the first term contributing to ${\mathbf W}^{-1}$ in Equation (\ref{eq:projectedellipse}). In other words, the shape derived from the inertia tensor of the two-dimensional mass distribution on the sky is generally not the same as the projected shape of the ellipsoid derived from the three-dimensional inertia tensor. This ambiguity arises because of our simplistic assumption of a top-hat radial light distribution, in one case implicitly imposed on the two-dimensional, and in the other on the three-dimensional galaxy model. We choose to implement the latter approach as it is more versatile with respect to future implementations of the galaxy shape model, e.g. the introduction of a misalignment between the major axes of dark matter halo and galaxy.

Using the definitions given in Section \ref{sec:Emodels}, the eigendecomposition of ${\mathbf M}$ can be written as
\eqa{
\label{eq:eigendecomp}
\mathbf{M} = \mathbf{V}\; \mathbf{D}\; \mathbf{V}^\tau, ~\mbox{where}~ D_{\alpha\beta} = \delta_{\alpha\beta} \omega_\beta^2\,; \mathbf{V} = \bc{ \vek{s}_1, \vek{s}_2,  \vek{s}_3 }\!.
}
Inserting these expressions into Equation (\ref{eq:projectedellipse}), and identifying $\alpha^2 = \vek{n}^\tau\, \mathbf{M}^{-1}\, \vek{n}$ as well as $\vek{k}=\mathbf{M}^{-1}\, \vek{n}$, it is straightforward to derive Equations (\ref{eq:ellipseprojection1}) and (\ref{eq:ellipseprojection2}).

\section{Extracting the intrinsic polarisation dispersion from weak lensing shear catalogues}
\label{app:cosmosshapes}

\subsection{Fisher measurement noise correction}
\label{app:noise}

To estimate the contribution by measurement noise to the observed polarisation dispersion, we follow the method proposed by \citet{leauthaud07}. Galaxy images are modelled as a bivariate Gaussian,
\eq{
\label{eq:gaussmodel}
G(\vek{x}) = \frac{S}{2 \pi \sqrt{ \det \mathbf{Q} }}\; \exp \bc{- \frac{1}{2} (\vek{x}-\vek{m})^\tau\, \mathbf{Q}^{-1}\, (\vek{x}-\vek{m})},
}
where $S$ denotes the flux, $\vek{m}$ the position of the image centre, and $\mathbf{Q}$ the symmetric second-order brightness tensor. These six parameters will in the following be collected into a six-dimensional parameter vector $\vek{p}$. 

The model $G(\vek{x})$ can be fit to the actual light distribution of a galaxy image. To avoid working at the image level, one resorts to the Fisher matrix which is given by the expectation value of the Hessian of the log-likelihood corresponding to this fit. This Fisher matrix is given by \citep{leauthaud07}
\eq{
\label{eq:noisefisher}
F_{\mu\nu} = \frac{1}{\sigma_N^2} \sum_i \frac{\partial G(\vek{x}_i)}{\partial p_\mu}\; \frac{\partial G(\vek{x}_i)}{\partial p_\nu}\;,
}
where $\sigma_N$ is the noise per image pixel\footnote{Note that our analysis is based on the \citet{schrabback09} reduction of the COSMOS data using the LANCZOS3 drizzle kernel and the native pixel scale $0.05^{\prime\prime}$, which minimises noise correlations between pixels.}. The sum runs over all pixels covered by the image, which in our implementation comprises the pixels within twice the half-light radius. Note that the derivatives in Equation (\ref{eq:noisefisher}) are readily computed analytically.

Assuming that the image centre is at $\vek{m}=0$, the only ingredients needed to compute the Fisher matrix are the flux $S$, the noise level $\sigma_N$ which can be inferred from the \texttt{SExtractor} ${\rm FLUXERR\_AUTO}$ parameter, and the three components of the brightness tensor $\mathbf{Q}$. The shear catalogue that we are working with provides us only with the polarisation components $e_{1,2}$, i.e. the image size information contained in $\mathbf{Q}$ is not directly available.

We retrieve this information in an approximate fashion by calculating the quantity
\eq{
\label{eq:qsizeinfo}
s \equiv \frac{Q_{11} + Q_{22}}{2} = \frac{\int \dd^2 x\; K_{\rm KSB}(\vek{x})\; I(\vek{x})\; |\vek{x}|^2}{2\; \int \dd^2 x\; K_{\rm KSB}(\vek{x})\; I(\vek{x})}\;,
}
where $K_{\rm KSB}$ is the KSB weighting, which \citet{schrabback09} chose as a circular Gaussian with the half-light radius as width. Note that the introduction of a weight function in the brightness moments is essential, as otherwise $\mathbf{Q}$ would be dominated by noise in the outskirts of the image. We model the galaxy's light distribution $I(\vek{x})$ as a circular Sersic profile with a typical Sersic index between 1 and 4 and a scale radius given in terms of the half-light radius. Then the elements of the brightness tensor are given by
\eq{
\label{eq:qelements}
Q_{11}=s\,(1+e_1)\;; ~~~ Q_{12}=s\,e_2\;; ~~~ Q_{22}=s\,(1-e_1)\;.
}

The flux $S$ required by the model in Equation (\ref{eq:gaussmodel}) is not necessarily equal to the measured flux of the galaxy image, as it is the best-fit amplitude of the Gaussian model. These two quantities can differ substantially since the Gaussian model does not provide very accurate fits. \citet{leauthaud07} introduced an overall calibration factor to account for this and determined it via image simulations. Instead, we choose an analytical route and calculate a calibration factor for each galaxy individually as the best-fit amplitude of a radial Gaussian profile to a radial Sersic profile, multiplied by the KSB kernel (both depending on the half-light radius of the galaxy). Note that this ansatz again makes the simplifying assumption of circular galaxy images.

After performing these steps, the Fisher matrix is computed for each galaxy. The submatrix corresponding to the three $\mathbf{Q}$ elements of the inverse Fisher matrix yields an estimate of the covariance ${\rm Cov}(Q)$ of $Q_{11}$, $Q_{12}$, and $Q_{22}$. Due to taking the inverse of the full six-dimensional Fisher matrix, the covariance is marginalised over uncertainties in the centroid position and the flux. The measurement error on $|e|^2$ can then be derived via
\eq{
\label{eq:epsnoise}
\Delta |e|^2 = \sqrt{ \sum_{i,j=1}^3 \frac{\partial |e|^2}{\partial Q_i}\; \bb{{\rm Cov}(Q)}_{ij} \frac{\partial |e|^2}{\partial Q_j}}\;,
}
where we introduced combined indices $i,j \in \bc{11;12;22}$. The derivatives are obtained in analytic form by making use of the definition of the polarisation in terms of the brightness tensor \citep{bartelmann01}. The error according to Equation (\ref{eq:epsnoise}) is computed for every galaxy, averaged over the samples and bins under consideration, and subtracted from the dispersion of $e$ obtained via Equation (\ref{eq:sigmaeps}).

Our computation of the brightness tensor elements requires $|e| \leq 1$, but in practice this limit can be exceeded for estimators of $e$ due to noise. This happens for 84 galaxies in our sample (all but one with $F814W>24$), which we discard completely.

\subsection{Circularisation correction of the polarisation}
\label{app:ksbkernel}

To determine the effect of the circular Gaussian kernel included in the brightness moments in the KSB implementation of \citet{schrabback09} on the measurement of galaxy ellipticity, we resort again to analytic light distributions using Sersic profiles. Varying the image polarisation between zero and unity (assuming all isophotes have the same polarisation), we compute the circular half-light radius for each profile. The result is then fed into the Gaussian kernel $K_{\rm KSB}$, which, together with the light distribution, is used to calculate the second brightness moments. By means of the defining equation \citep{bartelmann01} the \lq observed\rq\ polarisation is derived from the brightness tensor.

\begin{figure}
\centering
\includegraphics[scale=.34,angle=270]{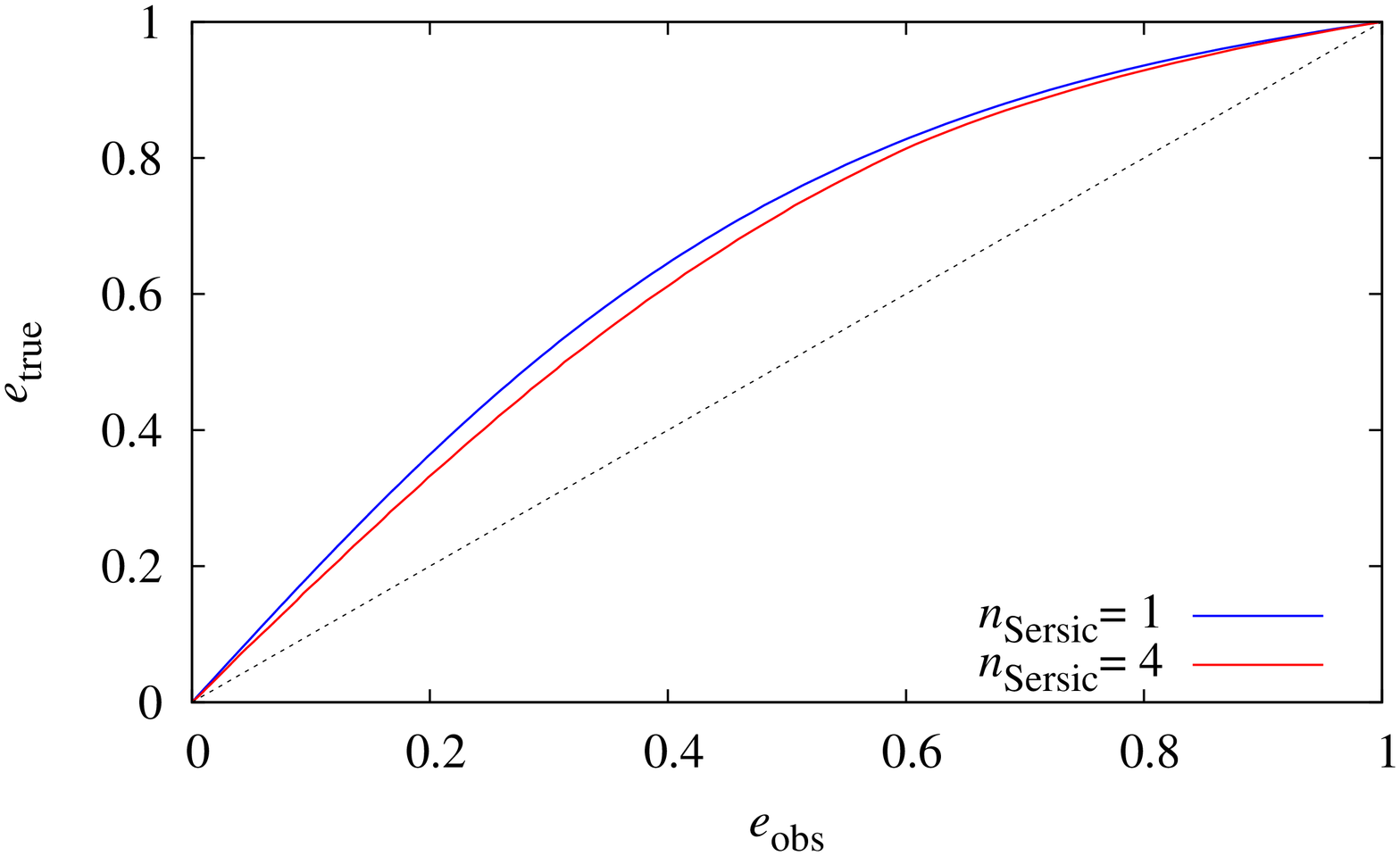}
\caption{Relation between the absolute value of the polarisation as measured by KSB, $e_{\rm obs}$, and the true polarisation $e_{\rm true}$. The blue (red) line results when assuming a galaxy light profile with Sersic index $n_{\rm Sersic}=1 (4)$. The black dotted line indicates a one-to-one relation.}
\label{fig:ksbeps}
\end{figure} 

In Fig.$\,$\ref{fig:ksbeps} the input (and hence true) absolute value of the polarisation in our computation, termed $e_{\rm true}$, is plotted against the resulting $e_{\rm obs}$ for a Sersic index of $n_{\rm Sersic}=1$, appropriate for late-type galaxies, and $n_{\rm Sersic}=4$, corresponding to a de Vaucouleurs profile typical of early-type galaxies. The observed polarisation is considerably smaller than the true galaxy polarisation, caused by the circular weighting entering the brightness moments. The dependence on the Sersic index is weak, so that we correct all galaxies according to a curve with $n_{\rm Sersic}=3.25$, which is close to the average of the two lines shown in the plot.

In our alternative approach to determine $\sigma_e$ the noise-corrected polarisations are subjected to the correction for the Gaussian kernel circularisation, based on the relation shown in Fig.$\,$\ref{fig:ksbeps}. This shifts the violet dotted line of Fig.$\,$\ref{fig:epserror} up to the black dotted line by about 0.15 in $\sigma_e$, independent of apparent magnitude by construction. Note that this correction is included in the \lq shear tensor\rq\ of the KSB formalism and thus automatically accounted for in the shear estimates which form the basis of our default approach.

Such a large correction may be regarded as a strong argument against employing weak lensing shear estimates for intrinsic galaxy shape measurements. However, it should be kept in mind that re-using these catalogues is not only convenient as all necessary steps to eliminate PSF effects and other systematics have already been performed, but weak lensing shape measurement methods might well be the only way to obtain reliable intrinsic shapes in the low signal-to-noise and small apparent size regime. Whether approaches other than KSB \citep[see e.g.][]{bridle09} are perhaps more suitable for this purpose remains the scope of future work.

\bibliographystyle{mn2e}

\label{lastpage}
\end{document}